\documentclass[12pt]{article}
\usepackage{t1enc}
\usepackage{amsmath}
\usepackage{amsthm}
\usepackage{latexsym}
\usepackage{natbib}
\usepackage{epsfig}
\usepackage{textcomp}
\usepackage{alltt}
\usepackage{graphicx}
\usepackage{rotating}
\usepackage{dcolumn}
\usepackage{amsfonts}
\usepackage{algorithm}
\usepackage[inline]{enumitem}

\begin{document}
\newcommand{\cip}{\perp\!\!\!\perp}
\newcommand{\nothere}[1]{}
\newcommand{\noi}{\noindent}
\newcommand{\mbf}[1]{\mbox{\boldmath $#1$}}
\newcommand{\cond}{\, |\,}
\newcommand{\hO}[2]{{\cal O}_{#1}^{#2}}
\newcommand{\hF}[2]{{\cal F}_{#1}^{#2}}
\newcommand{\tl}[1]{\tilde{\lambda}_{#1}^T}
\newcommand{\la}[2]{\lambda_{#1}^T(Z^{#2})}
\newcommand{\I}[1]{1_{(#1)}}
\newcommand{\cd}{\mbox{$\stackrel{\mbox{\tiny{\cal D}}}{\rightarrow}$}}
\newcommand{\cp}{\mbox{$\stackrel{\mbox{\tiny{p}}}{\rightarrow}$}}
\newcommand{\cas}{\mbox{$\stackrel{\mbox{\tiny{a.s.}}}{\rightarrow}$}}
\newcommand{\ld}{\mbox{$\; \stackrel{\mbox{\tiny{def}}}{=3D} \; $}}
\newcommand{\nk}{\mbox{$n \rightarrow \infty$}}
\newcommand{\con}{\mbox{$\rightarrow $}}
\newcommand{\dprime}{\mbox{$\prime \vspace{-1 mm} \prime$}}
\newcommand{\Borel}{\mbox{${\cal B}$}}
\newcommand{\bevis}{\mbox{$\underline{\em{Proof}}$}}
\newcommand{\Rd}[1]{\mbox{${\Re^{#1}}$}}
\newcommand{\il}[1]{{\int_{0}^{#1}}}
\newcommand{\pl}[1]{\mbox{\bf {\LARGE #1}}}
\newcommand{\expit}{\text{expit}}
\newcommand{\indep}{\rotatebox[origin=c]{90}{$\models$}}
\newcommand{\blind}{1}
\newcommand{\pr}{\text{pr}}
\newcommand{\var}{\text{var}}
\newcommand{\Bin}{\text{Bin}}
\newcommand{\Exp}{\text{Exp}}
\newcommand{\unif}{\text{unif}}
\newcommand{\logit}{\text{logit}}
\newcommand{\sign}{\text{sign}}
\newcommand{\support}{\text{support}}
\newcommand\norm[1]{\left\lVert#1\right\rVert}
\setlength{\bibsep}{0pt plus 0.3ex}

\newtheorem{theorem}{Theorem}
\newtheorem{lemma}{Lemma}
\newtheorem{prop}{Proposition}
\newtheorem{assumption}{Assumption}
\newtheorem{definition}{Definition}
\newtheorem*{remark}{Remark}
\newtheorem{corollary}{Corollary}
\newtheorem{example}{Example}

\parindent12pt

\begin{center}{\Large{\textbf{Uniformly valid confidence intervals for conditional treatment effects in misspecified high-dimensional models}}}
 \end{center}
 
 { 
\begin{center}
Oliver Dukes\\
\textit{Department of Applied Mathematics, Computer Science and Statistics}\\
\textit{Ghent University, Belgium}\\[2ex]
and Stijn Vansteelandt\\
\textit{Department of Applied Mathematics, Computer Science and Statistics}\\
\textit{Ghent University, Belgium}\\
\textit{and Department of Medical Statistics}\\
\textit{London School of Hygiene and Tropical Medicine, U.K.}\\[2ex]
email: oliver.dukes@ugent.be
\end{center}
}

\bigskip \setlength{\parindent}{0.3in} \setlength{\baselineskip}{24pt}

\begin{abstract}
Eliminating the effect of confounding in observational studies typically involves fitting a model for an outcome adjusted for covariates. When, as often, these covariates are high-dimensional, this necessitates the use of sparse estimators such as the Lasso, or other regularisation approaches. Na\"ive use of such estimators yields confidence intervals for the conditional  treatment effect parameter that are not \textit{uniformly valid}. Moreover, as the number of covariates grows with sample size, correctly specifying a model for the outcome is non-trivial. In this work, we deal with both of these concerns simultaneously, delivering confidence intervals for conditional treatment effects that are uniformly valid, regardless of whether the outcome model is correct. This is done by incorporating an additional model for the treatment-selection mechanism. When both models are correctly specified, we can weaken the standard conditions on model sparsity. Our procedure extends to multivariate treatment effect parameters and complex longitudinal settings.
\end{abstract}

\section{Introduction}

We focus on the problem of constructing confidence intervals for a low-dimensional component in a high-dimensional conditional mean model. In epidemiologic studies, this component may correspond to the effect of a discrete-valued exposure $A$ on an outcome $Y,$ conditional on a set of baseline covariates $L$. When the dimension of the covariates is large (relative to the sample size), data-adaptive model selection methods (such as the Lasso) are typically used to select a final regression model, on the basis of which inference on the conditional treatment effect is performed. However, standard inferential techniques ignore the additional uncertainty induced by the selection process. More seriously, they may also fail to be \textit{uniformly valid}; there may be no sample size at which a given procedure is guaranteed to attain its nominal coverage/size. In particular, the treatment effect estimator may have a complex, non-normal distribution (due to uncertainty in the model selection step), even when the sample size is very large. The series of models considered may also fail to contain the true model for the conditional mean of $Y$, given $A$ and $L$. 

Broadly speaking, there have been two main approaches for obtaining valid inferences on a low-dimensional parameter that depends on a high-dimensional regression adjustment. The first is based on doubly robust estimating equations \citep{robins_estimation_1994}, where a working model for the treatment-selection mechanism is postulated (a.k.a. the propensity score model), as well as for the outcome. Doubly robust estimators are unbiased when at least one of these working models is correctly specified. \citet{van_der_laan_targeted_2006} originally proposed combining this framework with flexible data-adaptive estimation of nuisance parameters. In the context of marginal treatment effects, \citet{farrell_robust_2015} shows that uniformly valid inferences can be obtained in high-dimensional settings by fitting both working models using the Lasso.  \citet{chernozhukov_double/debiased_2018} extend this work in many directions, for example allowing for a wide class of machine learning methods to estimate the nuisance parameters indexing the working models. If sample splitting is used, uniformly valid inferences are available under simple and generic conditions \citep{chernozhukov_double/debiased_2018}. Essentially, the predictions from each regression must converge to the truth, and the product of the $\ell_2$ norms of the prediction errors must shrink as $o_p(n^{-1/2})$. In high-dimensional parametric models, the latter condition requires that the product of the number of non-zero coefficients in each model must be small relative to the sample size. 

The second strand of work instead focuses on estimating the target parameter by `de-biasing' or `de-sparsifying' an initial Lasso-based estimate \citep{javanmard_confidence_2014,van_de_geer_asymptotically_2014,zhang_confidence_2014} or score equation \citep{chernozhukov_valid_2015,ning_general_2017}. Here, the bias in the penalised estimator is typically corrected via a single iteration of a Newton-Raphson style scheme. The bias correction term also depends on an additional regression adjustment, 
but one which does not necessarily correspond to a meaningful model for treatment-selection. Instead its role is purely to mitigate the bias incurred by estimating the parameter via the Lasso; after the de-biasing, (under certain conditions) the updated estimator is uniformly consistent and asymptotically normal. An advantage of this approach is its generality; in contrast, doubly robust estimators are only known to exist for a limited class of parameters. However, much stronger conditions are required on the sparsity of both the model for $Y$ and/or the de-biasing regression adjustment (unless certain tailored sparse estimators are used). In some settings, both approaches may coincide e.g. under the partially linear regression model.

In this work, we show how to construct confidence intervals for parameters in high-dimensional linear and log-linear regression models which are uniformly valid, regardless of whether the outcome model is correctly specified. We work instead under a correct model for the exposure, since in many epidemiologic studies, clinicians may have some knowledge on which variables affect the decision to give treatment and how they do so. Moreover, in complex settings (with time-varying treatments and covariates), coherent specification of multiple models for the outcome is non-trivial, so one typically prefers to do inference based on the propensity score \citep{robins_causal_1997}. Hence a different perspective is taken to recent research focusing on settings where a propensity score is misspecified or difficult to estimate well  \citep{athey_approximate_2018}. In linear models, our intervals are \textit{uniformly doubly robust}; valid if either a model for the exposure or outcome is correctly specified. We achieve this robustness by considering specific penalised estimators of the nuisance parameters.  In comparison, the inferences in \citet{farrell_robust_2015} and  \citet{chernozhukov_double/debiased_2018} are not generally robust to misspecification of either working model. When both models are correct (and a location-shift assumption holds), our proposal is valid under weakened conditions on model sparsity, without requiring sample-splitting. As we will discuss, our results here have wider ramifications for the use of machine learning algorithms in estimating causal effects. In the nonparametric setting, \citet{benkeser_doubly_2017} describe how to obtain doubly robust inference in settings where one of the data-adaptive estimators does not converge to the truth. Their focus is however on marginal treatment effects, and their results are not strictly applicable in high-dimensional settings, since they are developed under Donsker conditions which prohibit the dimension of the covariates growing with the sample size. We hence take an alternative approach.


Our work generalises the proposal of \cite{dukes_high-dimensional_2019}, where the focus was on hypothesis tests under the causal null hypothesis of no treatment effect. In comparison, we show how to construct uniformly valid confidence intervals and tests away from the null, describe a more general class of estimators for the nuisance parameters and extend the proposal to multivariate treatment effect parameters. 


\section{Proposal}\label{proposal}

\subsection{Model and motivation}

We will consider the model $\mathcal{M}$ defined by the restriction
\[g\{E(Y|A=a,L=l)\}-g\{E(Y|A=0,L=l)\}=\psi_0 a,\] where $g(\cdot)$ is a known link function and $\psi_0$ is an unknown parameter. For continuous $Y$, one might use the identity link $g(x)=x$ so that $\psi_0$ encodes the mean difference, or if $Y$ is constrained to only taking positive values, the log link $g(x)=\log(x)$ (so $\exp(\psi_0)$ is a ratio of expectations). These two choices of link function are the focus of this paper. Model $\mathcal{M}$ assumes that there is no treatment heterogeneity with respect to $L$ on the scale determined by the link function; we will weaken this restriction in Section \ref{extensions}. If one is willing to assume that $L$ is sufficient to adjust for confounding (along with the other standard conditions in the causal inference literature), then $\psi_0$ can be interpreted (either on the additive or multiplicative scale) as the average causal effect of removing a unit of treatment on the mean of $Y$, conditional on $L$.

Since $\psi_0$ can be expressed as a functional of conditional expectations, it is tempting to estimate it (and construct confidence intervals) based on postulating a parametric model $\mathcal{B}$ for the  conditional mean of the outcome. For example, consider the model $E(Y|A=0,L)=m(L;\beta_0)$, where $m(L;\beta)$ is a known function smooth in $\beta$ and $\beta_0$ is an unknown finite-dimensional parameter. 
Typically some dimension reduction is needed when the number of covariates is large relative to the sample size. Estimating $\beta_0$ via the Lasso \citep{tibshirani_regression_1996} or the Dantzig selector \citep{candes_dantzig_2007} is convenient because they enforce a sparse solution; components of the estimate of $\beta_0$ will likely be set to zero. Alternatively, these estimators could be an intermediate step for selecting covariates to be included in a final model. 

However, this raises two concerns. The first is that in finite samples, the distribution of a sparse estimator $\tilde{\beta}$ is typically complex \citep{knight_asymptotics_2000}. One cannot in general rule out the existence of covariates that weakly predict the outcome, but are strongly associated with the exposure, such that $\beta_0$ contains components that are close (but not equal) to zero. The estimator of these entries may be forced to zero in certain samples but not others, and the resulting estimator of $\psi_0$ based on $\tilde{\beta}$ will tend to inherit this non-regular behaviour. The consequence is that standard confidence intervals (based on the normal approximation) are not uniformly valid, in the sense that for any finite $n$, there exist parts of the parameter space for which the interval coverage may be poor \citep{leeb_model_2005}. The second concern is that the true model for $E(Y|A=0,L)$ may not be nested within the series of regressions considered during the selection process. When $L$ is high-dimensional, specification of a correct model for $Y$ is especially challenging, particularly in observational studies where the distribution of the covariates differs greatly between treatment groups. In this case, model $\mathcal{B}$ will tend to extrapolate to regions outside of the observed data range, and small changes in the model may greatly impact conclusions on the treatment effect. 

\subsection{Doubly robust scores for conditional treatment effects}

Although our focus is obtaining valid confidence intervals, for ease of exposition we will begin by considering the problem of testing the hypothesis $\psi=\psi_0$. In Section \ref{invert}, we will then link back to the construction of intervals. 

In order to construct our test, we will require two regression adjustments; the first is based on model $\mathcal{B}$ for the conditional mean of the outcome. The second is a model $\mathcal{A}$ for the conditional mean of the exposure (a.k.a. the propensity score when $A$ is binary); namely $E(A|L)=\pi(L;\gamma_0)$ where $\pi(L;\gamma)$ is smooth in $\gamma$ and $\gamma_0$ is an unknown finite-dimensional parameter. For binary $A$, one typically uses a logistic model e.g. $\pi(L;\gamma_0)=\expit(\gamma_0^TL)$. Our test statistic will then be based on the score
\[U(\psi,\eta)=\{A-\pi(L;\gamma)\}\{H(\psi)-m(L;\beta)\}\]
\citep{robins_estimating_1992}. Here, 
$\eta=(\gamma^T,\beta^T)^T$ and $H(\psi)=Y\expit(-\psi A)$ if $g(\cdot)$ is the log link; otherwise $H(\psi)=Y-\psi A$. In what follows, we will allow one of the models to be misspecified, such that $\gamma_0$ or $\beta_0$ no longer agrees with the truth. Even in that case, using the law of iterated expectation one can show that $E\{U(\psi_0,\eta_0)\}=0$ if either $E(Y|A=0,L)=m(L;\beta_0)$ or $E(A|L)=\pi(L;\gamma_0)$; this property of double robustness will be key for obtaining uniformly valid inference (even when model $\mathcal{B}$ is misspecified).


Once we have obtained estimates $\hat{\eta}$ of $\eta$, we can construct a test of $\psi=\psi_0$ based on the statistic:
\begin{align}\label{score}
T_n(\psi_0,\hat{\eta})=\hat{V}^{-1/2}\frac{1}{\sqrt{n}}\sum^n_{i=1}U_i\{\psi_0,\hat{\eta}(\psi_0)\}
\end{align}
where $\hat{V}$ is the empirical estimate of the variance of $U(\psi_0,\hat{\eta})$ and $\hat{\eta}(\psi_0)$ makes explicit that $\eta_0$ is estimated at the fixed value $\psi_0$. In the following section, we will propose specific estimators $\hat{\eta}(\psi_0)$ of $\eta_0$ under the assumption that $\psi=\psi_0$. In Section \ref{asymptotics} we discuss the conditions under which the statistic (\ref{score}) is uniformly asymptotically normal. 

\subsection{Estimation of the nuisance parameter $\eta$}\label{np_motiv}

For the moment, we will work under the model $\mathcal{M}\cap\mathcal{A}$ - in other words, we will assume that in addition to the semiparametric model $\mathcal{M}$, the propensity score model for the exposure holds. We will postulate a logistic model for the exposure $\pi(L;\gamma_0)=\expit(\gamma_0^TL)$; because our setting is high-dimensional, we will estimate $\gamma_0$ by fitting this model with a Lasso penalty e.g. we solve the minimization problem:
\begin{align}\label{gamma_est}
\hat{\gamma}=\arg\min_{\gamma}\frac{1}{n}\sum^{n}_{i=1}\log\{1+\exp(\gamma^TL_i)\}-A_i(\gamma^TL_i)+\lambda_\gamma||\gamma||_1
\end{align}
where $\lambda_\gamma >0$ is the penalty parameter and $\norm{.}_1$ denotes the $\ell_1$ norm. To improve finite sample performance, in practice we recommend refitting this model adjusted for the selected covariates using maximum likelihood. Unfortunately, in the asymptotic distribution of the score $U(\psi,\eta)$, terms like  
\begin{align}\label{grad_2}
\frac{1}{n}\sum^n_{i=1}\frac{\partial}{\partial \gamma}U_i\{\psi_0, \hat{\eta}(
\psi_0)\}\sqrt{n}(\gamma_0-\hat{\gamma})
\end{align}
are problematic for inference, because the distribution of $\gamma_0-\hat{\gamma}$ can be complex and difficult to approximate well.

We therefore recommend constructing an estimator $\hat{\beta}(\psi_0)$ at which:
\begin{align}\label{bridge}
0=\frac{1}{n}\sum^n_{i=1}\frac{\partial}{\partial \gamma}U_i\{\psi_0, \hat{\eta}(\psi_0)\}+\lambda_\beta\delta|\hat{\beta}(\psi_0)|^{\delta-1}\circ\sign\{\hat{\beta}(\psi_0)\}
\end{align}
\citep{fu_penalized_2003}. Here $\lambda_\beta>0$, $\delta\geq 1$, $\circ$ is the Hadamard product operator and for a vector $a\in\mathbb{R}^{p}$, $\sign(a)$ is a vector of elements  $\sign(a_j)$ (for $j=1,...,p$). Also, $\delta|\beta|^{\delta-1}\circ\sign(\beta)$ refers to the partial derivative of $||\beta||^\delta_\delta$ with respect to $\beta$; the $\ell_\delta$ norm is defined as $||a||_\delta\equiv\bigg(\sum^p_{i=1}|a_i|^\delta\bigg)^{1/\delta}$. Then we define $\beta_0$ as the solution to the population analogue of the above estimating equations (without the penalty term) that corresponds with the truth when model $\mathcal{B}$ is correct. Above, the gradient $\partial U(\psi_0,\hat{\eta}(\psi_0)/\partial \gamma$ is used as an \textit{estimating function} for $\beta_0$, so as to ensure that $\sum^n_{i=1} \partial U_i(\psi_0,\hat{\eta}(\psi_0)/\partial \gamma$ is close to zero at the estimator of the nuisance parameter. Specifically, letting $w(L;\gamma)=\expit(\gamma^TL)\{1-\expit(\gamma^TL)\}$, we are proposing to estimate $\beta_0$ as the solution to
\begin{align}\label{beta_est}
0=&-\frac{1}{n}\sum^n_{i=1}w(L_i;\hat{\gamma})\{H_i(\psi_0)-m(L_i;\beta)\}L_i+\lambda_\beta\delta|\beta|^{\delta-1}\circ\sign(\beta).
\end{align}
We will let $\delta$ converge to 1, in order that a sparse solution will be returned and the procedure can be implemented using software for the Lasso.  

Estimating $\beta_0$ as described above ensures that
\begin{align*}
\norm{\frac{1}{n}\sum^n_{i=1}\frac{\partial}{\partial \gamma}U_i\{\psi_0, \hat{\eta}(\psi_0)\}}_\infty &= ||\lambda_\beta\delta|\hat{\beta}(\psi_0)|^{\delta-1}\circ\sign\{\hat{\beta}(\psi_0)\}||_\infty \leq \delta\lambda_\beta
\end{align*}
since $||\delta|\hat{\beta}(\psi_0)|^{\delta-1}\circ\sign\{\hat{\beta}(\psi_0)\}||_\infty \leq 1$ for $\delta\to 1 +$. So for penalty terms satisfying the standard condition that $\lambda_\beta=O(\sqrt{\log (p\lor n)/n})$  (where $a \lor b$ denotes the maximum of $a$ and $b$) and assuming $\log (p\lor n)=o(n)$, it follows that the $\ell_\infty$ norm of the gradient term asymptotically goes to zero. 

More generally, we require estimators of $\beta$ that have the property that
\begin{align}\label{rn}
\norm{\frac{1}{n}\sum^n_{i=1}\frac{\partial}{\partial \gamma}U_i\{\psi_0, \hat{\eta}(\psi_0)\}}_\infty\leq C \lambda_\beta
\end{align}
where $C$ is a constant and $\lambda_\beta$ is a positive tuning parameter that converges to zero as $n\to \infty$. Since this idea is developed from the theory of bias-reduced doubly robust estimation \citep{vermeulen_bias-reduced_2015}, we describe this as the `high-dimensional bias reduction' property. Whilst in the previous work, the procedure was motivated by preventing the inflation of asymptotic bias under misspecification of one or both working models, here we use it to minimise the impact of using a non-regular estimator of $\gamma_0$. 

In what follows (and in the proofs in Appendix \ref{appA}), we will focus on estimating $\beta_0$ via a bridge penalty, letting $\delta\to 1 +$. However, there exist other estimators which also obtain the `high-dimensional bias reduction' property. If one is happy to postulate a linear outcome model for $\mathcal{B}$ e.g. $m(L;\beta)=\beta^TL$, then one can use a Dantzig-based estimator of $\beta$:
\begin{align}\label{dantzig}
\hat{\beta}(\psi_0)=&\arg\min_{\beta} \norm{\beta}_1\quad \textrm{s.t.}\quad\norm{\frac{1}{n}\sum^n_{i=1}w(L_i;\hat{\gamma})\{H_i(\psi_0)-\beta^TL_i\}L_i}_{\infty}\leq \lambda_\beta
\end{align}
Similar to (\ref{beta_est}), this also sets the relevant gradient approximately to zero. It is a generalization of the proposal of \citet{ning_general_2017}, where they use a similar approach in fitting a model for the exposure. If we are not willing to assume ultra-sparsity in a model for $Y$, then one can also adapt the methodology of \citet{zhu_significance_2016}, basing estimation of $\beta_0$ on the solution path of the linear program:
\begin{align*}
\hat{\beta}(\psi_0)=\arg\min \norm{\beta}_1\quad\textrm{s.t.}&\quad\norm{\frac{1}{n}\sum^n_{i=1}w(L_i;\hat{\gamma})\{H_i(\psi_0)-\beta^TL_i\}L_i}_{\infty}\leq \lambda_\beta\\
& \quad \norm{\sum^n_{i=1}\{H_i(\psi_0)-\beta^TL_i\}}_{\infty}\leq \kappa\\
& \quad \frac{1}{n}\sum^n_{i=1}w(L_i;\hat{\gamma})\{H_i(\psi_0)-\beta^TL_i\}H_i(\psi_0)\geq \bar{\kappa}
\end{align*}
where $\kappa$ and $\bar{\kappa}$ are positive tuning parameters. Compared with (\ref{dantzig}), this approach includes two extra constraints; these allow for the outcome model $\mathcal{B}$ to be dense by controlling the relevant remainder terms in the distribution of $\psi_0$. 
We do not consider either of the estimation approaches described in this paragraph any further in this paper, since it's currently unclear how feasible they are for non-linear models. 

We close this section by noting that in low-dimensional settings, when estimating the variance of the doubly robust estimator, gradient terms like $(\ref{grad_2})$ are often ignored. The motivation is that if an efficient estimator of $\gamma_0$ is used, then pretending the propensity score is known will generally yield conservative inferences when model $\mathcal{B}$ is misspecified \citep{robins_estimating_1992}. However, as far as we are aware, this result currently has no analogue in the high-dimensional setting (where sparse estimation of $\gamma_0$ is required) and thus using estimators of $\beta_0$ that lack the `high-dimensional bias reduction' property may not be guaranteed to yield intervals that exceed their nominal coverage level. Moreover, we would expect reduced bias for an estimator of $\psi_0$ based on $\hat{\beta}(\psi_0)$ when model $\mathcal{B}$ is grossly misspecified, relative to using an arbitrary sparse estimator of $\beta$ (given the results in \citet{vermeulen_bias-reduced_2015}). This is because the weights $w(L;\gamma)$ will make the resulting estimator less prone to extrapolating outside of the observed data range (since regions of low overlap in $L$ will be given weights close to zero).

\subsection{Inverting the score test}\label{invert}

Plugging in estimates $\hat{\eta}$ of $\eta_0$ and scaling $U\{\psi_0,\hat{\eta}(\psi_0)\}$, one can now obtain a statistic $T_n\{\psi_0,\hat{\eta}(\psi_0)\}$.  
Given the conditions discussed in the following section, we will argue that by the form of the score equation and the choice of estimators of $\eta_0$, it follows that under model $\mathcal{M}\cap\mathcal{A}$, $T^2_n\{\psi_0,\hat{\eta}(\psi_0)\}\overset{p}{\to}\chi^2_1 $ where $\chi^2_1$ denotes a chi-squared distribution on 1 degree of freedom. Hence $T_n\{\psi_0,\hat{\eta}(\psi_0)\}$ can be used to straightforwardly test the hypothesis that $\psi=\psi_0$.

We can adapt this reasoning to construct a $(1-\alpha)100\%$ confidence interval for $\psi_0$ as
\begin{align}\label{CI}
[\hat{l}_s,\hat{u}_s]=&
\left(\psi_0:\left[\frac{1}{n}\sum^n_{i=1}U_i\{\psi_0,\hat{\eta}(\psi_0)\}\right]^2-\frac{\chi^2_1(\alpha)}{n}\hat{V}\leq0 \right)
\end{align}
where $\chi^2_1(\alpha)$ is the critical value of $\chi^2_1$ corresponding to the significance level $\alpha$. In practice, we will search over a grid of values of $\psi$ in order to find the values $l_s$ and $u_s$ that satisfy the above inequality; note that $\beta_0$ will be re-estimated under each considered $\psi$. 
Furthermore, using the same reasoning, we can obtain a point estimate of $\psi_0$ as $\hat{\psi}=\arg\min_\psi T^2_n\{\psi,\hat{\eta}(\psi)\}$.

In the following section, we will discuss the theoretical properties of the intervals given above, and indicate the specific benefits of inverting the score test as proposed.

\section{Asymptotic properties}\label{asymptotics}

Let $\mathcal{P}'$ be the class of laws that obey the intersection submodel $\mathcal{M}\cap\mathcal{A}$; then we are interested in convergence under a sequence of laws $P_n\in \mathcal{P}'$. We will allow for $p$ to increase with $n$, and for the values of the population parameters $\psi_0$, $\gamma_0$ and $\beta_0$ to depend on $n$, and hence also models $\mathcal{A}$ and $\mathcal{B}$ (although the notation will be suppressed for convenience). Note that at a given $n$, we will assume the existence of a sparse parameter $\beta_0$ that is the solution to the unpenalised population analogue of the equations in (\ref{bridge}). We use $\mathbb{P}_{P_n}[]$ to denote a probability taken with respect to the local data generating process $P_n$. Let us define the active set of variables as $S_\gamma=\support(\gamma_0)$ and $S_\beta=\support(\beta_0)$. Furthermore, let $s_\gamma$ denote the cardinality $|S_\gamma|$ and likewise $s_\beta=|S_\beta|$. We will use the following result to show that $\hat{l}_s$ and $\hat{u}_s$ in (\ref{CI}) form a uniformly valid confidence interval; the proofs of all results are left to Appendix \ref{appA}. 

\begin{theorem}\label{theorem1}
If, in addition to Assumptions \ref{moment_CI}-\ref{cons_p} in Appendix \ref{appA},\\
\begin{enumerate*}[label=(\roman*)]
\item $(s^2_\gamma+s^2_\beta)\log^2 (p\lor n)=o(n)$ \label{US_sum_CI}\\
\end{enumerate*}
holds, then - using estimators $\hat{\gamma}$ and $\hat{\beta}(\psi_0)$ defined in (\ref{gamma_est}) and (\ref{bridge}) - we have
\begin{align}\label{UC_result}
\lim_{n\to \infty}\sup_{P_n\in\mathcal{P}'}\big|\mathbb{P}_{P_n}\left(\psi_0\in[\hat{l}_s,\hat{u}_s]\right)-(1-\alpha)\big|= 0
\end{align}
under model $\mathcal{M}\cap\mathcal{A}$. 
\end{theorem}
This result shows that under `ultra-sparse' regimes ($s_\gamma<<\sqrt{n}$ and $s_\beta<<\sqrt{n}$), one can construct a uniformly valid interval for $\psi_0$ without requiring a correct outcome model $\mathcal{B}$. Fitting the working model for $Y$ in the specific way proposed above helps to correct for the regularisation bias incurred via the sparse estimate $\hat{\gamma}$, similar to the literature on de-biasing the Lasso \citep{belloni_post-selection_2016,ning_general_2017}. Indeed, the ultra-sparsity condition in that literature is standard if one restricts to estimation via the Lasso or the Dantzig selector. The key difference is that we do not require a correct model for $\mathcal{B}$. 

Stronger results are available on robustness to misspecification when the working model for the outcome is linear:
\begin{corollary}\label{corr1}
Suppose that $m(L;\beta)$ is linear with respect to $\beta$. Then under the same conditions as Theorem \ref{theorem1}, the confidence interval 
$[\hat{l}_s,\hat{u}_s]$ is uniformly valid as in (\ref{UC_result}) under the union model $\mathcal{M}\cap(\mathcal{A}\cup\mathcal{B})$. 
\end{corollary}
In this case, the resulting intervals are \textit{uniformly doubly robust}, in the sense that they should contain the true parameter with probability determined by the nominal $\alpha$-level when either model $\mathcal{A}$ or $\mathcal{B}$ is correct, uniformly over the parameter space. Stronger conditions on $s_\gamma$ and $s_\beta$ are not required. In principle, uniformly doubly robust confidence intervals could be constructed when the outcome model is non-linear. However, this is challenging computationally as estimating $\gamma_0$ now requires weights dependent on $\hat{\beta}(\psi_0)$ such that iteration is required. This would have to be done over all values of $\psi$ considered in solving (\ref{CI}).

If all models are correct and a particular location-shift condition holds, then one can weaken the corresponding assumptions on model sparsity.
\begin{theorem}\label{theorem2}
Let us restrict our consideration to the class of laws $\mathcal{P}$ that obey the intersection sub-model  $\mathcal{M}\cap\mathcal{A}\cap\mathcal{B}$. We also suppose that\\
\begin{enumerate*}[label=(\roman*)]\addtocounter{enumi}{1}
\item $(s_\gamma+s_\beta)\log (p\lor n)=o(n)$ \label{S_sum_CI}\\
\item $(s_\gamma s^*)\log^2 (p\lor n)=o(n)$ \label{prod_CI}\\
\item $H(\psi_0)\indep A|L$ \label{l_shift}\\
\end{enumerate*} 
hold, where $s^*=s_\gamma \lor s_\beta$.  
Then if $m(L;\beta_0)$ is linear in $\beta_0$, under Assumptions \ref{moment_CI}, \ref{concentration}, \ref{cons_p}, \ref{strong_cons_CI} and \ref{bounded_res_CI} in Appendix \ref{appA} and the conditions \ref{S_sum_CI}-\ref{l_shift}, the confidence interval  $[\hat{l}_s,\hat{u}_s]$ is uniformly valid as in (\ref{UC_result}). For general models for $m(L;\beta_0)$, the same result holds if $\beta_0$ is estimated from a subsample of the data separate to the one used to construct the interval, without requiring condition \ref{l_shift}.
\end{theorem}

For $m(L;\beta_0)=\beta_0^TL$, when both models $\mathcal{A}$ and $\mathcal{B}$ are correct (in addition to model $\mathcal{M}$) and model $\mathcal{A}$ is ultra-sparse, one can allow for model $\mathcal{B}$ to be dense (and vice versa). Hence we describe our confidence intervals as \textit{sparsity adaptive}. Condition \ref{l_shift} would hold under the semiparametric location-shift model 
\begin{align}\label{ls_model}
Y=\psi_0 A+\epsilon
\end{align}
where $\epsilon \indep A|L$. 
If $L$ is sufficient to adjust for confounding of the effect of $A$ on $Y$, we can rephrase model (\ref{ls_model}) as a \textit{linear structural distribution model} \citep{robins_causal_1997}.

With non-linear $m(L;\beta_0)$, we revert to sample splitting to relax the sparsity assumptions, although we conjecture that uniform validity under weakened conditions is also possible here. This is partly because results in \citet{dukes_high-dimensional_2019} imply that confidence intervals obtained via our procedure without weighting are valid under the intersection submodel if conditions \ref{S_sum_CI}, \ref{prod_CI} and \ref{l_shift} hold (see also the corollary below). It also follows from the proofs in Appendix \ref{appA} that without sample splitting, so long as all models are correct, ultra-sparsity is only required in model $\mathcal{B}$ a.k.a we require $s_\gamma \log (p\lor n)=o(n)$ and $s^2_\beta \log^2 (p\lor n)=o(n)$. 

When $H(\psi_0)=Y-\psi_0 A$, \citet{chernozhukov_double/debiased_2018} arrive at conditions \ref{S_sum_CI} and \ref{prod_CI} without requiring \ref{l_shift} via the use of sample splitting. Moreover, as long as their recommended `cross-fitting' scheme is used, asymptotically there should be little or no efficiency loss. Nevertheless, the benefits of sample splitting are currently only apparent when estimators of both $\beta_0$ and $\gamma_0$ converge to the truth, and it may become infeasible with limited sample sizes or  in more complex causal inference settings (see section \ref{extensions}). When their score equations are not linear in the target parameter, the regularity conditions in \citet{chernozhukov_double/debiased_2018} are less intuitive even when combined with sample splitting, whereas the confidence intervals proposed above are valid under simpler conditions regardless of whether $H(\psi_0)$ is linear in $\psi_0$, largely by virtue of inverting a score test. 

The sparsity adaptivity property does not appear to be available for Wald-based intervals. To see why, note that for testing $\psi=\psi_0$, a Wald test based on the score $U(\psi,\eta)$ would require fitting a model for $E(Y|A,L)$ and discarding an initial estimate $\check{\psi}$ of $\psi$. The resulting estimates $\check{\beta}$ of $\beta$ are then dependent on $(A_i)^n_{i=1}$, whereas $\hat{\beta}(\psi_0)$ (as proposed above) can only depend on the exposure data via the transformed outcome $H(\psi_0)$. In the proof of Theorem \ref{theorem2}, we exploit this property - combined with the conditional independence of  $H(\psi_0)$ and $A$ from \ref{l_shift} - to emulate settings where $\beta$ is estimated in a separate sample. We note that the construction of intervals in high-dimensional models by inverting tests has been described before e.g. in \citet{chernozhukov_valid_2015}, although there it was used more as a pedagogic device  rather than specifically to weaken sparsity assumptions.

In a final corollary, we indicate the consequences of these results for a broader class of machine learning algorithms. This result is implied by the proofs of Theorem \ref{theorem2}. 
\begin{corollary}
Suppose that we obtain unweighted estimators $\hat{\pi}(L)$ and $\hat{m}(L;\psi_0)$ of $E(A|L)=\pi(L)$ and $E(Y|A=0,L)=m(L)$ respectively via machine learning estimators and we repeat the above steps, inverting the test statistic $T_n\{\psi_0,\hat{\pi},\hat{m}(\psi_0)\}$ (replacing $T_n\{\psi_0,\hat{\eta}(\psi_0)\}$) to obtain an interval $[\check{l}_s,\check{u}_s]$. Furthermore, we will assume the estimators satisfy $n^{-1}\sum^n_{i=1}\{\hat{\pi}(L_i)-\pi(L_i)\}^2=o_{P_n}(1)$, $n^{-1}\sum^n_{i=1}\{\hat{m}(L_i;\psi_0)-m(L_i)\}^2=o_{P_n}(1)$, and 
\begin{align*}
&\left[n^{-1}\sum^n_{i=1}\{\hat{\pi}(L_i)-\pi(L_i)\}^2\right]^{1/2}\left[n^{-1}\sum^n_{i=1}\{\hat{m}(L_i;\psi_0)-m(L_i)\}^2\right]^{1/2}=o_{P_n}(n^{-1/2}).
\end{align*}
Then so long as Assumption \ref{moment_CI} and \ref{l_shift} holds, under the class of laws $\mathcal{P}$, $[\check{l}_s,\check{u}_s]$ is a uniformly valid interval as defined above.
\end{corollary}
By inverting a score test and utilising the location-shift condition, one can use arbitrary machine learning estimators in constructing the interval  without having to either use sample splitting or invoke strong Donsker-type conditions. For a discussion of which estimators meet these conditions, see \citet{chernozhukov_double/debiased_2018}. We emphasise this only applies so long as both estimators converge to the truth.

\section{Simulation study}

In each of the 1,000 simulations comprising our Experiment 1, we created a dataset with $n=200$ observations. We generated the covariates $L^*$ from a multivariate normal distribution $\mathcal{N}(0,\Sigma)$, where $\Sigma$ is a Toeplitz matrix with $\Sigma_{j,k}=2^{-|j-k|}$; then we created $L$ by including an additional column for the intercept. The dimension of $L$ was $p=200$. Further, $A$ was a Bernoulli random variable with conditional expectation $E(A|L)=\expit(\gamma^T_0L)$ and $Y$ was generated from the normal distribution, $\mathcal{N}(0.3 A+\beta_0^TL,1)$, where $\beta_0=\tau(-1,1,-1,2^{\rho},...,(p-2)^{\rho})$. As in \citet{farrell_robust_2015}, we used $\tau$ to vary the signal strength (with 1 indicating a stronger signal and 0.4 a weaker one) and $\rho$ to control sparsity (with 2 indicating a sparser model and 0.5 more dense). In this and all subsequent experiments, $\gamma_0=1,-1,1,-2^{-2},...,(p-2)^{-2}$. 
In Experiment 2, we created the covariates $X_1=|\log(5+L^*_1)|$, $X_2=L^*_2\exp(L^*_1)$ and $X_3=-(L^*_2+L^*_3)^2$. A matrix $X$ was created by binding $X_1$, $X_2$ and $X_3$, along with columns $4$ to $p$ of $L^*$ (plus a column corresponding to the intercept). Then we generated $N(0.3 A+\bar{\beta}_0^TX,1)$, where $\bar{\beta}_0$ was the same as $\beta_0$ except the leading three entries were equal to 1. Experiment 3 was similar to Experiment 1, except now $Y\sim N\{0.3 A+\beta_0^TL,\sigma(A,L)\}$, where $\sigma(A_i,L_i)=\{ n^{-1}\sum^n_i (0.3 A_i+\beta_0^TL_i)^2\}^{-1/2}(0.3 A_i+\beta_0^TL_i)$. Each experiment was repeated with $p=250$, varying $\tau$ and $\rho$. 

We first considered a na\"ive post-selection approach, where $Y$ was regressed on $A$ and $L$ using a Lasso-penalty (selected via 20-fold cross-validation), forcing the exposure into the model. The final model was then refit, adjusted for $A$ and the selected covariates, yielding the estimate $\hat{\psi}_{OLS}$. We compared this with the `post-double selection' method, as described in \citet{belloni_inference_2014}, and implemented using the `hdm' package in R \citep{chernozhukov_hdm:_2016}, as well the `partialling out' method in the same package (yielding the estimators $\hat{\psi}_{PDS}$ and $\hat{\psi}_{PO}$ respectively). 
Both approaches were implemented using the penalties selected by the `hdm' package; we also present results for post-double selection and partialling out using penalties instead obtained via cross-validation (let $\hat{\psi}_{PDS-CV}$ and $\hat{\psi}_{PO-CV}$ denote the respective estimators). For the estimators $\hat{\psi}_{OLS}$, $\hat{\psi}_{PDS}$, $\hat{\psi}_{PO}$,  $\hat{\psi}_{PDS-CV}$ and $\hat{\psi}_{PO-CV}$, we used `model-based' variance estimators given by the software. 

For our approach, each working model was adjusted for $L$. Lasso penalties for the working models for exposure and outcome were both selected using 20-fold cross-validation, choosing $\lambda_\gamma$ as the value that minimised the expected cross-validated error (and likewise for $\lambda_\beta$). In the case of the outcome model, cross-validation was done under the null $\psi=0$, which we would expect to generally yield anti-conservative penalties. If too many covariates were selected such that refitting model $\mathcal{A}$ using maximum likelihood failed, a small increment was added to $\lambda_\gamma$. In Experiment 1, both models were correctly specified, whereas in Experiment 2, only the logistic model for the exposure was correct. In Experiment 3, both models were correct again; however, condition \ref{l_shift} in Theorem \ref{theorem2} was violated due to the dependence of the residual variance on the exposure. A point estimate of the treatment effect (denoted by $\hat{\psi}_{HDBR}$) and confidence intervals were otherwise obtained as in Section \ref{invert}. To compare the efficiency of $\hat{\psi}_{HDBR}$ with the other estimators in the `MSE' column of the tables, we evaluated the sample standard error of the score function $U\{\psi, \hat{\eta}(\psi)\}$, with $\psi$ held fixed at the true value.

\begin{table}[htbp]
\centering
\caption{Simulation results from Experiment 1 ($n=200$). Estimators considered (Est); Monte Carlo bias multiplied by 10 (Bias); Monte Carlo standard deviation multiplied by 10 (MCSD); Mean  standard error multiplied by 10 (MSE); coverage probability multiplied by 100 (Cov).}
\resizebox{\textwidth}{!}{\begin{tabular}{llllllllll}
  \hline
   &  & \multicolumn{4}{c}{$p=200$} & \multicolumn{4}{c}{\textbf{$p=250$}}\\

 $\rho,\tau$ & Est & Bias & MCSD & MSE & Cov  & Bias & MCSD & MSE & Cov \\
\hline
2,1 & $\hat{\psi}_{OLS}$ & -0$\cdot$78 & 2 & 1$\cdot$6 & 85$\cdot$4 & -0$\cdot$94 & 1$\cdot$9 & 1$\cdot$5 & 83$\cdot$8 \\ 
   & $\hat{\psi}_{PDS}$ & -0$\cdot$5 & 1$\cdot$9 & 1$\cdot$7 & 90$\cdot$6 & -0$\cdot$6 & 1$\cdot$9 & 1$\cdot$7 & 91$\cdot$3 \\ 
   & $\hat{\psi}_{PO}$ & -0$\cdot$82 & 1$\cdot$7 & 1$\cdot$6 & 91$\cdot$1 & -0$\cdot$91 & 1$\cdot$7 & 1$\cdot$6 & 91$\cdot$3 \\ 
   & $\hat{\psi}_{PDS-CV}$ & -0$\cdot$73 & 1$\cdot$9 & 1$\cdot$8 & 91$\cdot$9 & -0$\cdot$81 & 1$\cdot$8 & 1$\cdot$8 & 92$\cdot$1 \\ 
   & $\hat{\psi}_{PO-CV}$ & -0$\cdot$91 & 1$\cdot$8 & 1$\cdot$6 & 88 & -1$\cdot$05 & 1$\cdot$7 & 1$\cdot$6 & 86$\cdot$3 \\ 
   & $\hat{\psi}_{HDBR}$ & -0$\cdot$72 & 2$\cdot$1 & 1$\cdot$9 & 92$\cdot$2 & -0$\cdot$86 & 2$\cdot$1 & 1$\cdot$9 & 90$\cdot$7 \\ 
  0$\cdot$5,1 & $\hat{\psi}_{OLS}$ & -1$\cdot$14 & 3$\cdot$1 & 2 & 78$\cdot$7 & -1$\cdot$69 & 3$\cdot$2 & 2$\cdot$1 & 73$\cdot$8 \\ 
   & $\hat{\psi}_{PDS}$ & -5$\cdot$71 & 3 & 2$\cdot$8 & 44 & -5$\cdot$81 & 3$\cdot$1 & 2$\cdot$8 & 43$\cdot$6 \\ 
   & $\hat{\psi}_{PO}$ & -5$\cdot$71 & 2$\cdot$9 & 2$\cdot$7 & 44$\cdot$8 & -5$\cdot$82 & 3 & 2$\cdot$8 & 43 \\ 
   & $\hat{\psi}_{PDS-CV}$ & -1$\cdot$18 & 2$\cdot$4 & 2$\cdot$3 & 90$\cdot$2 & -1$\cdot$28 & 2$\cdot$6 & 2$\cdot$4 & 89$\cdot$1 \\ 
   & $\hat{\psi}_{PO-CV}$ & -1$\cdot$73 & 1$\cdot$7 & 1$\cdot$5 & 67$\cdot$2 & -1$\cdot$76 & 2 & 1$\cdot$7 & 69 \\ 
   & $\hat{\psi}_{HDBR}$ & -1$\cdot$13 & 2$\cdot$6 & 2$\cdot$3 & 92$\cdot$3 & -1$\cdot$5 & 2$\cdot$9 & 2$\cdot$6 & 91$\cdot$2 \\ 
  2,0$\cdot$4 & $\hat{\psi}_{OLS}$ & -1$\cdot$56 & 2$\cdot$1 & 1$\cdot$5 & 74$\cdot$7 & -1$\cdot$76 & 2$\cdot$2 & 1$\cdot$5 & 68$\cdot$4 \\ 
   & $\hat{\psi}_{PDS}$ & -2$\cdot$78 & 1$\cdot$6 & 1$\cdot$6 & 59$\cdot$6 & -2$\cdot$9 & 1$\cdot$7 & 1$\cdot$7 & 57$\cdot$4 \\ 
   & $\hat{\psi}_{PO}$ & -2$\cdot$78 & 1$\cdot$6 & 1$\cdot$7 & 60$\cdot$9 & -2$\cdot$91 & 1$\cdot$7 & 1$\cdot$7 & 57$\cdot$6 \\ 
   & $\hat{\psi}_{PDS-CV}$ & -0$\cdot$72 & 1$\cdot$9 & 1$\cdot$8 & 92$\cdot$2 & -0$\cdot$77 & 2$\cdot$1 & 1$\cdot$9 & 88$\cdot$2 \\ 
   & $\hat{\psi}_{PO-CV}$ & -0$\cdot$84 & 1$\cdot$9 & 1$\cdot$7 & 89$\cdot$7 & -0$\cdot$85 & 2 & 1$\cdot$8 & 86$\cdot$7 \\ 
   & $\hat{\psi}_{HDBR}$ & -0$\cdot$66 & 2$\cdot$1 & 2 & 92$\cdot$6 & -0$\cdot$68 & 2$\cdot$1 & 2$\cdot$1 & 92$\cdot$1 \\ 
  0$\cdot$5,0$\cdot$4 & $\hat{\psi}_{OLS}$ & -2$\cdot$18 & 2$\cdot$3 & 1$\cdot$7 & 68$\cdot$9 & -2$\cdot$34 & 2$\cdot$2 & 1$\cdot$7 & 65$\cdot$1 \\ 
   & $\hat{\psi}_{PDS}$ & -2$\cdot$95 & 1$\cdot$9 & 1$\cdot$8 & 62$\cdot$9 & -3 & 1$\cdot$8 & 1$\cdot$8 & 61$\cdot$1 \\ 
   & $\hat{\psi}_{PO}$ & -2$\cdot$95 & 1$\cdot$9 & 1$\cdot$8 & 63$\cdot$7 & -3$\cdot$01 & 1$\cdot$8 & 1$\cdot$8 & 61$\cdot$7 \\ 
   & $\hat{\psi}_{PDS-CV}$ & -1$\cdot$08 & 2$\cdot$2 & 2 & 89$\cdot$6 & -1$\cdot$2 & 2$\cdot$1 & 2 & 87$\cdot$5 \\ 
   & $\hat{\psi}_{PO-CV}$ & -1$\cdot$16 & 2$\cdot$1 & 1$\cdot$9 & 85$\cdot$3 & -1$\cdot$31 & 2$\cdot$1 & 1$\cdot$9 & 83$\cdot$8 \\ 
   & $\hat{\psi}_{HDBR}$ & -1$\cdot$02 & 2$\cdot$3 & 2$\cdot$2 & 91$\cdot$4 & -1$\cdot$06 & 2$\cdot$4 & 2$\cdot$3 & 92$\cdot$7 \\ 
\hline
\end{tabular}}
\label{exp1tab}
\end{table}

\begin{table}[htbp]
\centering
\caption{Simulation results from Experiment 2 ($n=200$). Estimators considered (Est); Monte Carlo bias multiplied by 10 (Bias); Monte Carlo standard deviation multiplied by 10 (MCSD); Mean  standard error multiplied by 10 (MSE); coverage probability multiplied by 100 (Cov)}
\resizebox{\textwidth}{!}{\begin{tabular}{llllllllll}
  \hline
   &  & \multicolumn{4}{c}{$p=200$} & \multicolumn{4}{c}{\textbf{$p=250$}}\\

 $\rho,\tau$ & Est & Bias & MCSD & MSE & Cov  & Bias & MCSD & MSE & Cov \\
\hline
2,1 & $\hat{\psi}_{OLS}$ & -1$\cdot$5 & 3$\cdot$8 & 2$\cdot$7 & 82$\cdot$2 & -1$\cdot$43 & 4 & 2$\cdot$6 & 80$\cdot$5 \\ 
   & $\hat{\psi}_{PDS}$ & -3$\cdot$6 & 2$\cdot$9 & 2$\cdot$9 & 79$\cdot$5 & -3$\cdot$6 & 3 & 2$\cdot$9 & 78$\cdot$9 \\ 
   & $\hat{\psi}_{PO}$ & -3$\cdot$6 & 2$\cdot$9 & 3$\cdot$1 & 84$\cdot$3 & -3$\cdot$6 & 3 & 3$\cdot$1 & 83$\cdot$6 \\ 
   & $\hat{\psi}_{PDS-CV}$ & -3$\cdot$49 & 3$\cdot$1 & 3$\cdot$2 & 81$\cdot$3 & -3$\cdot$58 & 3$\cdot$2 & 3$\cdot$2 & 81 \\ 
   & $\hat{\psi}_{PO-CV}$ & -3$\cdot$46 & 3 & 3 & 79$\cdot$6 & -3$\cdot$56 & 3$\cdot$1 & 3 & 80$\cdot$1 \\ 
   & $\hat{\psi}_{HDBR}$ & 0$\cdot$3 & 3 & 2$\cdot$9 & 95 & 0$\cdot$24 & 3$\cdot$1 & 3 & 93$\cdot$6 \\ 
  0$\cdot$5,1 & $\hat{\psi}_{OLS}$ & -2$\cdot$76 & 6$\cdot$4 & 5$\cdot$3 & 87$\cdot$5 & -1$\cdot$99 & 6$\cdot$3 & 5$\cdot$3 & 89$\cdot$4 \\ 
   & $\hat{\psi}_{PDS}$ & -7$\cdot$17 & 5$\cdot$9 & 5$\cdot$9 & 78$\cdot$2 & -6$\cdot$81 & 5$\cdot$8 & 5$\cdot$9 & 81$\cdot$2 \\ 
   & $\hat{\psi}_{PO}$ & -7$\cdot$17 & 5$\cdot$9 & 6$\cdot$4 & 84 & -6$\cdot$81 & 5$\cdot$8 & 6$\cdot$5 & 85$\cdot$6 \\ 
   & $\hat{\psi}_{PDS-CV}$ & -6$\cdot$85 & 6$\cdot$3 & 6$\cdot$7 & 84$\cdot$8 & -6$\cdot$4 & 6$\cdot$6 & 6$\cdot$9 & 85$\cdot$5 \\ 
   & $\hat{\psi}_{PO-CV}$ & -6$\cdot$84 & 6$\cdot$3 & 6$\cdot$6 & 83$\cdot$7 & -6$\cdot$31 & 6$\cdot$6 & 6$\cdot$7 & 84$\cdot$7 \\ 
   & $\hat{\psi}_{HDBR}$ & -0$\cdot$17 & 5$\cdot$9 & 6 & 95$\cdot$1 & 0$\cdot$23 & 6$\cdot$3 & 6$\cdot$2 & 95$\cdot$4 \\ 
  2,0$\cdot$4 & $\hat{\psi}_{OLS}$ & 0$\cdot$16 & 2$\cdot$3 & 1$\cdot$7 & 85 & 0$\cdot$18 & 2$\cdot$3 & 1$\cdot$7 & 86$\cdot$5 \\ 
   & $\hat{\psi}_{PDS}$ & -1$\cdot$41 & 2 & 2 & 89 & -1$\cdot$46 & 2$\cdot$1 & 2 & 88$\cdot$2 \\ 
   & $\hat{\psi}_{PO}$ & -1$\cdot$41 & 2 & 2$\cdot$1 & 90$\cdot$8 & -1$\cdot$47 & 2$\cdot$1 & 2$\cdot$1 & 89$\cdot$9 \\ 
   & $\hat{\psi}_{PDS-CV}$ & -1$\cdot$51 & 2$\cdot$1 & 2$\cdot$1 & 88$\cdot$9 & -1$\cdot$58 & 2$\cdot$2 & 2$\cdot$1 & 87$\cdot$2 \\ 
   & $\hat{\psi}_{PO-CV}$ & -1$\cdot$55 & 2$\cdot$1 & 2$\cdot$1 & 87$\cdot$1 & -1$\cdot$61 & 2$\cdot$2 & 2$\cdot$1 & 85$\cdot$8 \\ 
   & $\hat{\psi}_{HDBR}$ & 0$\cdot$06 & 2$\cdot$4 & 2$\cdot$3 & 94$\cdot$8 & 0$\cdot$1 & 2$\cdot$5 & 2$\cdot$3 & 93$\cdot$3 \\ 
  0$\cdot$5,0$\cdot$4 & $\hat{\psi}_{OLS}$ & -0$\cdot$83 & 3 & 2$\cdot$5 & 89$\cdot$5 & -0$\cdot$77 & 2$\cdot$9 & 2$\cdot$5 & 91$\cdot$2 \\ 
   & $\hat{\psi}_{PDS}$ & -2$\cdot$73 & 2$\cdot$8 & 2$\cdot$9 & 85$\cdot$4 & -2$\cdot$84 & 2$\cdot$9 & 2$\cdot$9 & 83$\cdot$6 \\ 
   & $\hat{\psi}_{PO}$ & -2$\cdot$73 & 2$\cdot$8 & 3$\cdot$1 & 88$\cdot$8 & -2$\cdot$84 & 2$\cdot$9 & 3$\cdot$1 & 87 \\ 
   & $\hat{\psi}_{PDS-CV}$ & -2$\cdot$72 & 3$\cdot$1 & 3$\cdot$2 & 87$\cdot$8 & -2$\cdot$87 & 3$\cdot$3 & 3$\cdot$3 & 86$\cdot$9 \\ 
   & $\hat{\psi}_{PO-CV}$ & -2$\cdot$74 & 3 & 3$\cdot$1 & 86$\cdot$7 & -2$\cdot$9 & 3$\cdot$2 & 3$\cdot$2 & 85 \\ 
   & $\hat{\psi}_{HDBR}$ & 0$\cdot$09 & 3$\cdot$3 & 3$\cdot$2 & 94$\cdot$4 & 0$\cdot$04 & 3$\cdot$4 & 3$\cdot$2 & 94$\cdot$9 \\ 
\hline
\end{tabular}}
\label{exp2tab}
\end{table}

\begin{table}[htbp]
\centering
\caption{Simulation results from Experiment 3 ($n=200$). Estimators considered (Est); Monte Carlo bias multiplied by 10 (Bias); Monte Carlo standard deviation multiplied by 10 (MCSD); Mean estimated standard error multiplied by 10 (MSE); coverage probability multiplied by 100 (Cov).}
\resizebox{\textwidth}{!}{\begin{tabular}{llllllllll}
  \hline
   &  & \multicolumn{4}{c}{$p=200$} & \multicolumn{4}{c}{\textbf{$p=250$}}\\

 $\rho,\tau$ & Est & Bias & MCSD & MSE & Cov  & Bias & MCSD & MSE & Cov \\
\hline
2,1 & $\hat{\psi}_{OLS}$ & -0$\cdot$87 & 1$\cdot$7 & 1$\cdot$6 & 89$\cdot$8 & -0$\cdot$95 & 1$\cdot$8 & 1$\cdot$5 & 84$\cdot$9 \\ 
   & $\hat{\psi}_{PDS}$ & -1$\cdot$09 & 2$\cdot$3 & 1$\cdot$5 & 83$\cdot$7 & -1$\cdot$18 & 2$\cdot$4 & 1$\cdot$5 & 81$\cdot$3 \\ 
   & $\hat{\psi}_{PO}$ & -1$\cdot$36 & 2$\cdot$1 & 1$\cdot$6 & 85$\cdot$6 & -1$\cdot$45 & 2$\cdot$2 & 1$\cdot$6 & 83$\cdot$8 \\ 
   & $\hat{\psi}_{PDS-CV}$ & -0$\cdot$81 & 1$\cdot$7 & 1$\cdot$8 & 94 & -0$\cdot$88 & 1$\cdot$9 & 1$\cdot$8 & 90$\cdot$7 \\ 
   & $\hat{\psi}_{PO-CV}$ & -1 & 1$\cdot$6 & 1$\cdot$6 & 88$\cdot$5 & -1$\cdot$04 & 1$\cdot$8 & 1$\cdot$6 & 85$\cdot$7 \\ 
   & $\hat{\psi}_{HDBR}$ & -0$\cdot$89 & 1$\cdot$8 & 1$\cdot$7 & 89$\cdot$9 & -0$\cdot$88 & 1$\cdot$8 & 1$\cdot$7 & 88$\cdot$7 \\ 
  0$\cdot$5,1 & $\hat{\psi}_{OLS}$ & -1$\cdot$26 & 3$\cdot$1 & 2 & 78$\cdot$5 & -1$\cdot$76 & 3$\cdot$2 & 2$\cdot$1 & 72$\cdot$6 \\ 
   & $\hat{\psi}_{PDS}$ & -5$\cdot$75 & 3$\cdot$1 & 2$\cdot$7 & 41$\cdot$5 & -6$\cdot$07 & 3 & 2$\cdot$7 & 39$\cdot$8 \\ 
   & $\hat{\psi}_{PO}$ & -5$\cdot$76 & 2$\cdot$9 & 2$\cdot$7 & 43$\cdot$7 & -6$\cdot$07 & 2$\cdot$9 & 2$\cdot$8 & 40$\cdot$5 \\ 
   & $\hat{\psi}_{PDS-CV}$ & -1$\cdot$24 & 2$\cdot$3 & 2$\cdot$3 & 91$\cdot$3 & -1$\cdot$43 & 2$\cdot$5 & 2$\cdot$4 & 88$\cdot$5 \\ 
   & $\hat{\psi}_{PO-CV}$ & -1$\cdot$73 & 1$\cdot$7 & 1$\cdot$5 & 67$\cdot$1 & -1$\cdot$88 & 1$\cdot$9 & 1$\cdot$6 & 63$\cdot$5 \\ 
   & $\hat{\psi}_{HDBR}$ & -1$\cdot$21 & 2$\cdot$5 & 2 & 91 & -1$\cdot$49 & 2$\cdot$7 & 2$\cdot$3 & 91$\cdot$1 \\ 
  2,0$\cdot$4 & $\hat{\psi}_{OLS}$ & -1$\cdot$59 & 2$\cdot$3 & 1$\cdot$5 & 69$\cdot$6 & -1$\cdot$59 & 2$\cdot$3 & 1$\cdot$5 & 70$\cdot$7 \\ 
   & $\hat{\psi}_{PDS}$ & -2$\cdot$9 & 1$\cdot$7 & 1$\cdot$6 & 54$\cdot$4 & -2$\cdot$84 & 1$\cdot$7 & 1$\cdot$6 & 57 \\ 
   & $\hat{\psi}_{PO}$ & -2$\cdot$91 & 1$\cdot$6 & 1$\cdot$7 & 57$\cdot$5 & -2$\cdot$85 & 1$\cdot$7 & 1$\cdot$7 & 59$\cdot$1 \\ 
   & $\hat{\psi}_{PDS-CV}$ & -0$\cdot$69 & 2 & 1$\cdot$8 & 90$\cdot$9 & -0$\cdot$79 & 2$\cdot$1 & 1$\cdot$8 & 90$\cdot$1 \\ 
   & $\hat{\psi}_{PO-CV}$ & -0$\cdot$77 & 1$\cdot$9 & 1$\cdot$8 & 89$\cdot$3 & -0$\cdot$91 & 2 & 1$\cdot$8 & 87$\cdot$6 \\ 
   & $\hat{\psi}_{HDBR}$ & -0$\cdot$6 & 2$\cdot$1 & 2$\cdot$1 & 92$\cdot$9 & -0$\cdot$7 & 2$\cdot$2 & 2$\cdot$1 & 91$\cdot$1 \\ 
  0$\cdot$5,0$\cdot$4 & $\hat{\psi}_{OLS}$ & -2$\cdot$12 & 2$\cdot$3 & 1$\cdot$7 & 66$\cdot$4 & -2$\cdot$31 & 2$\cdot$2 & 1$\cdot$7 & 67 \\ 
   & $\hat{\psi}_{PDS}$ & -3$\cdot$02 & 1$\cdot$8 & 1$\cdot$8 & 59$\cdot$3 & -3$\cdot$1 & 1$\cdot$8 & 1$\cdot$8 & 58$\cdot$4 \\ 
   & $\hat{\psi}_{PO}$ & -3$\cdot$03 & 1$\cdot$8 & 1$\cdot$8 & 60$\cdot$5 & -3$\cdot$1 & 1$\cdot$8 & 1$\cdot$8 & 60$\cdot$4 \\ 
   & $\hat{\psi}_{PDS-CV}$ & -1$\cdot$12 & 2$\cdot$1 & 2 & 88$\cdot$2 & -1$\cdot$26 & 2$\cdot$2 & 2 & 87$\cdot$3 \\ 
   & $\hat{\psi}_{PO-CV}$ & -1$\cdot$22 & 2 & 1$\cdot$9 & 84$\cdot$7 & -1$\cdot$31 & 2$\cdot$1 & 1$\cdot$9 & 83$\cdot$3 \\ 
   & $\hat{\psi}_{HDBR}$ & -1$\cdot$02 & 2$\cdot$2 & 2$\cdot$1 & 89$\cdot$6 & -1$\cdot$17 & 2$\cdot$4 & 2$\cdot$2 & 89$\cdot$2 \\ 
\hline
\end{tabular}}
\label{exp3tab}
\end{table}

Results for the three experiments are given in Tables \ref{exp1tab}-\ref{exp3tab}. 
They indicate that even in highly sparse, strong signal settings where the model for $Y$ is correctly specified, the na\"ive approach still has a large bias with standard errors that do not adequately reflect the uncertainty induced by the Lasso procedure. The post-double selection and partialling out methods performed better in this case, but often failed to attain the nominal coverage level either under denser models or when the signal was weaker. Part of the poorer performance was due to the choice of penalty terms; use of cross-validation improved results considerably. This was particularly true for post-double selection, which performed surprisingly well in Experiment 3, despite the fact that the variance estimator used is not generally robust to heteroscedastic errors. Results for these methods were comparable or worse under misspecification of the outcome model, indicating that performance is very sensitive to the data generating mechanism. In contrast, we saw that our proposed confidence intervals came close to attaining their nominal coverage across the majority of settings. They performed poorest in dense settings where the signal was strong, as well as when errors were heteroscedastic (as predicted by the theory). However, they still generally improved upon alternatives. 
Experiments 1-3 were repeated with $n=p=400$, where superior coverage was seen across all settings (see Appendix \ref{appB}).


\section{Extensions}\label{extensions}
\subsection{Effect heterogeneity and categorical exposures}\label{effect_hetero}


Suppose that interest lies in the exposure effect parameter $\psi=(\psi^{(1)},\psi^{(2)})^T$ indexing the semiparametric model $\mathcal{M}_{int}$ defined by
\[g\{E(Y|A=a,L=l)\}-g\{E(Y|A=0,L=l)\}=\psi_0^{(1)} a+\psi_0^{(2)}az.\] 
Here, $Z$ is a scalar component of $L$.  We can now redefine $H(\psi_0)$, such that $H(\psi_0)=Y-\psi_0^{(1)} A-\psi_0^{(2)}AZ$ when $g(\cdot)$ is the identity link and $H(\psi_0)=Y\exp(-\psi_0^{(1)} A-\psi_0^{(2)}AZ)$ when $g(\cdot)$  is the log link. 

Then because $\psi_0$ is now two-dimensional, we recommend the use of different nuisance parameters in the first and second doubly robust estimating functions for $\psi_0$. In particular, consider the following equations:
\[U(\psi,\eta)=\begin{pmatrix}
\{A-\pi(L;\gamma)\}\{H(\psi)-m(L;\beta^{(1)})\}\\Z\{A-\pi(L;\gamma)\}\{H(\psi)-m(L;\beta^{(2)})\}
\end{pmatrix}=
\begin{pmatrix}
U^{(1)}(\psi,\eta)\\U^{(2)}(\psi,\eta)
\end{pmatrix}.
\]
Following the reasoning in Section \ref{np_motiv}, we use the gradient of $U^{(1)}(\psi,\eta)$ with respect to $\gamma$ as an estimating function for $\beta_0^{(1)}$ (and likewise for $\beta_0^{(2)}$). So we now estimate $\beta_0^{(1)}$ and $\beta_0^{(2)}$ as the solutions respectively  to 
\begin{align}
0&=-\sum^n_{i=1}w(L_i;\hat{\gamma})\{H_i(\psi_0)-m(L_i;\beta^{(1)})\}L_i+\lambda_{\beta^{(1)}}\delta|\beta^{(1)}|^{\delta-1}\circ\sign(\beta^{(1)})\label{eh_ee1}\\
0&=-\sum^n_{i=1}w(L_i;\hat{\gamma})\{H_i(\psi_0)-m(L_i;\beta^{(2)})\}L_iZ_i+\lambda_{\beta^{(2)}}\delta|\beta^{(2)}|^{\delta-1}\circ\sign(\beta^{(2)})\label{eh_ee2}.
\end{align}
By allowing $\beta$ to take on different values in the different estimating equations for $\psi$, we enable the targeting of the nuisance parameter estimates towards the different parameters of interest. We now search over a two-dimensional space for $\psi$, estimating $\beta_0^{(1)}$ and $\beta_0^{(2)}$ at each value considered. Then $T_n\{\psi_0,\hat{\eta}(\psi_0)\}$ can be compared to a $\chi^2_2$ distribution, and inverting the test statistic yields a confidence interval for $\psi_0^{(1)}$ and $\psi_0^{(2)}$ that is uniformly valid under model $\mathcal{M}_{int}\cap\mathcal{A}$ (following the proof of Theorem \ref{theorem1}). 
Our proposal contrasts with other approaches for estimating interaction effects in high-dimensional models (e.g. \citet{belloni_inference_2014} and \citet{chernozhukov_double/debiased_2018}), where separate models are postulated for $E(A|L)$ and $E(AZ|L)$. In contrast, we only require a single exposure model $\mathcal{A}$, which is also computationally more efficient, not to mention more robust if the outcome model is misspecified. For instance, if linear models are postulated for both $E(A|L)$ and $E(AZ|L)$, then they cannot generally both be correct.  

Similarly, if $A$ is an exposure with three categories (taking values 0,1 or 2), then the semiparametric model $\mathcal{M}_{cat}$ is now
\[g\{E(Y|A=a,L=l)\}-g\{E(Y|A=0,L=l)\}=\psi^{(1)}_0 a_1+\psi^{(2)}_0a_2\] 
where $A_1=1$ if $A=1$ and 0 otherwise, and $A_2=1$ if $A=2$ and 0 otherwise. Also, $H(\psi_0)$ is redefined accordingly e.g. $H(\psi_0)=Y-\psi_0^{(1)} A_1-\psi_0^{(2)}A_2$ when $g(\cdot)$ is the identity link. Then the estimating functions for $\psi^{(1)}$ and $\psi^{(2)}$ are
\[U(\psi,\eta)=\begin{pmatrix}
\{A_1-\pi(L;\gamma^{(1)})\}\{H(\psi)-m(L;\beta^{(1)})\}\\ \{A_2-\pi(L;\gamma^{(2)})\}\{H(\psi)-m(L;\beta^{(2)})\}
\end{pmatrix}\]
where $\pi(L;\gamma_0^{(1)})$ is a model postulated for $P(A=1|L)\equiv P(A_1=1|L)$ and $\pi(L;\gamma_0^{(2)})$ is a model postulated for $P(A=2|L)\equiv P(A_2=1|L)$. Postulating a multinomial logistic model $\mathcal{A}$ for $A$, the parameters $\gamma_0^{(1)}$ and $\gamma_0^{(2)}$ can be estimated efficiently via the Group Lasso (see \citet{farrell_robust_2015} for theoretical guarantees). Then $\beta_0^{(1)}$ and $\beta_0^{(2)}$ can be estimated as in (\ref{eh_ee1}) and (\ref{eh_ee2}) (except $\hat{\beta}^{(1)}$ will only depend on $\hat{\gamma}^{(1)}$, and likewise $\hat{\beta}^{(2)}$ depends only on $\hat{\gamma}^{(2)}$). 

A drawback however is that even when a linear model is postulated for the conditional mean of $Y$, the confidence intervals are valid under model $\mathcal{M}_{int}\cap\mathcal{A}$ (or $\mathcal{M}_{cat}\cap\mathcal{A}$) but are not uniformly doubly robust. This is because, in the case of effect modification, estimating $\gamma$ as in (\ref{gamma_est}) will not set $\sum^n_{i=1}\partial U^{(2)}(\psi_0,\hat{\eta})/\partial \beta^{(2)}$ approximately to zero. This can be addressed by estimating separate parameters $\gamma^{(1)}$ and $\gamma^{(2)}$ in the same way as is done for $\beta$ above. In the case of categorical exposures, one must fit separate logistic models for $P(A=1|L)$ and $P(A=2|L)$ rather than using a Group Lasso approach. 


\subsection{Controlled direct effects}

Finally, we will briefly consider a study with a variable $A_1$ measured at baseline along with an accompanying collection of variables $L_1$. These may confound the association between $A_1$ and an end of study outcome $Y$. In addition, we also measure a post-baseline variable $A_2$, which may be influenced by $L_2$ (covariates measured after baseline but prior to $A_2$) along with $L_1$ and $A_1$. Interest is in the causal effect of $A_2$ on $Y$, but also the effect of $A_1$ on $Y$  if $A_2$ were fixed at zero. Under extended structural assumptions (see below), the latter corresponds to  \textit{controlled direct effect} of $A_1$ on $Y$ \citep{robins_identifiability_1992}. One may consider $A_2$ as a mediator of the effect of $A_1$, or as a subsequent measurement of a time-varying exposure $A$.  Furthermore, we allow $L_2$ to depend on the baseline variable $A_1$, which means that standard regression-based approaches do not generally encode the controlled direct effect of $A_1$ \citep{robins_new_1986}. 

For time $t=1,2$, we will postulate the model $\mathcal{M}$:
\begin{align*}
&g\{E(Y|A_t=a_t,\bar{A}_{t-1}=\bar{a}_{t-1},\bar{L}_{t}=\bar{l}_{t})\}\\&-g\{E(Y|A_t=0,\bar{A}_{t-1}=\bar{a}_{t-1},\bar{L}_{t}=\bar{l}_{t})\}=\psi^{(t)}_0a_t;
\end{align*}
here, $\bar{L}_t$ and $\bar{A}_{t-1}$ denote the covariate and the exposure history up to time $t$, with $A_0=\emptyset$. The model imposes the restriction that the effect $\psi^{(t)}_0$ of $A_t$ on $Y$ does not depend on $\bar{L}_t$ and $\bar{A}_{t-1}$ at each $t$. In order to give either of the above contrasts a causal interpretation, a sequential `no unmeasured confounding' assumption is required. Namely, $\bar{L}_t$ and $\bar{A}_{t-1}$ must suffice to adjust for confounding between $A_t$ and $Y$. Therefore in the two time-point setting, under this condition $\psi^{(2)}_0$ encodes the (conditional) causal effect of $A_2$ and $Y$ and $\psi^{(1)}_0$ the controlled direct effect of $A_1$ on $Y$ (fixing $A_2$ at zero).

To construct tests and confidence intervals for these parameters, we will postulate the models $\mathcal{A}_t$ for $E(A_t|\bar{A}_{t-1},\bar{L}_t)$ e.g. $E(A_t|\bar{A}_{t-1},\bar{L}_t)=\pi_t(\bar{A}_{t-1},\bar{L}_t;\gamma_0^{(t)})$, and $\mathcal{B}_t$ for $E(Y|A_t=0,\bar{A}_{t-1},\bar{L}_t)$ e.g. $E(Y|A_t=0,\bar{A}_{t-1},\bar{L}_t)=m_t(\bar{A}_{t-1},\bar{L}_t;\beta_0^{(t)})$. Inference may then be obtained via the estimating functions
\[U(\psi,\eta)=\begin{pmatrix}
\{A_2-\pi_2(\bar{L}_2,A_1;\gamma^{(2)})\}\{H_2(\psi^{(2)})-m_2(\bar{L}_2,A_1;\beta^{(2)})\}\\\{A_1-\pi_1(L_1;\gamma^{(1)})\}\{H_1(\psi)-m_1(L_1;\beta^{(1)})\}
\end{pmatrix}\]
where  $\psi=(\psi^{(1)},\psi^{(2)})^t$ and $\eta=(\gamma^{(1)^T},\gamma^{(2)^T},\beta^{(1)^T},\beta^{(2)^T})^T$. For an identity link, $H_2(\psi^{(2)})=Y-\psi^{(2)}A_2$ and $H_1(\psi)=Y-\psi^{(2)}A_2-\psi^{(1)}A_1$; otherwise $H_2(\psi^{(2)})=Y\exp(-\psi^{(2)}A_2)$ and $H_1(\psi)=Y\exp(-\psi^{(2)}A_2-\psi^{(1)}A_1)$. 

\section{Discussion}

In this paper, we have described how to obtain uniformly valid confidence intervals for the conditional treatment effect parameters in high-dimensional linear and log-linear models. We have unified and generalised the existing doubly robust and de-biasing approaches: unified, since our proposal adapts to the sparsity conditions in each literature, depending on the modelling assumptions one is willing to make; and generalised, since we allow for misspecification and more general model choices. This allows us to extend our work to a wide selection of problems (like estimating controlled direct effects), where one does not wish to rely on an outcome model being correct. Unfortunately, it's currently unclear how our proposal could be extended to the conditional causal odds ratio, since no doubly robust estimator of this parameter currently exists under the union model $\mathcal{M}\cap(\mathcal{A}\cup\mathcal{B})$ when $g(\cdot)$ is the logit link; the same applies to the hazard ratio.

In future work, we will look at incorporating more general machine learning methods in the construction of confidence intervals. In general, under model $\mathcal{M}\cap\mathcal{A}$, stronger rate conditions on the estimators are required than under the intersection sub-model, which are currently available for a limited selection of estimators. These include the Lasso, post-Lasso and more recently, deep neural networks \citep{farrell_deep_2018}. In fact, if conditions equivalent to those in Appendix \ref{appA} are met, it follows that deep neural networks could be substituted for the Lasso for estimating the propensity score. It is an open question whether the high-dimensional bias reduction property exists for a more general class of machine learning estimators; such a development would be useful when the conditional expectation $E(Y|A=a,L)$ is difficult to estimate well. 

\section*{Acknowledgments}

The first author is supported by a PhD grant from the Research Foundation - Flanders (FWO).

\appendix
\bigskip \setlength{\parindent}{0.3in} \setlength{\baselineskip}{24pt}
\numberwithin{equation}{section}\section{Appendix A}\label{appA} 

In this Appendix, we give proofs of the main results in Section \ref{asymptotics} of the main paper. Beginning with some notation, we use $\mathbb{E}_{P_n}[]$ for taking expectation w.r.t. the local data generating process (DGP), whereas $\mathbb{E}_n[]$ refers to sample expectations. Similarly, $\mathbb{P}_{P_n}[]$ and  $\var_{P_n}[]$ denote probabilities and variances taken w.r.t. the local DGP respectively. In certain places, we will use the notation $\hat{\beta}(\psi_0,\hat{\gamma})$, in order to make explicit the dependence of the estimator of $\beta$ on the estimated weights. The gradients $\partial U_i\{\psi_0, \hat{\eta}(\psi_0)\}/\partial \eta$ and $\partial U_i\{\psi_0, \eta_0\}/\partial \eta$ are viewed as row vectors.

In the proofs that follow, we will make the following assumptions: 
\begin{assumption}\label{moment_CI}
(Moment conditions). For some constants $0<c<C<\infty$ and $4<r<\infty$,
\begin{enumerate}[label=(\roman*)]
\item $\mathbb{E}_{P_n}[\{H(\psi_0)-m(L;\beta_0)\}^4|A,L]<C$ with probability approaching 1. \label{m1}
\item $\mathbb{E}_{P_n}[|H(\psi_0)-m(L;\beta_0)|^{r}]<C$. \label{m2}
\item $c<\mathbb{E}_{P_n}[\{A-\pi(L;\gamma_0)\}^2|L]$ and $c<\mathbb{E}_{P_n}[\{H(\psi_0)-m(L;\beta_0)\}^2|A,L]$ with probability approaching 1. \label{m3}
\end{enumerate}
\end{assumption}
\begin{assumption}\label{concentration}
(Concentration bound). Let $d(L;\beta_0)=\partial m(L;\beta_0)/\partial \beta$; then
\begin{align*}
\norm{\frac{1}{\sqrt{n}}\sum^n_{i=1}\{A_i-\pi(L_i;\gamma_0)\}d(L_i;\beta_0)}_\infty&=O_{P_n}(\sqrt{\log (p\lor n)}).
\end{align*}
\end{assumption}
\begin{assumption}\label{cons_e}
(Rates of convergence of the parameter estimates). For a given sequence $P_n$, the estimators $\hat{\gamma}$ and $\hat{\beta}(\psi_0)$ satisfy:
\begin{enumerate}[label=(\roman*)]
\item $\norm{\hat{\gamma}-\gamma_0}_1=O_{P_n}(s_\gamma\sqrt{\log(p \lor n)/n})$. \label{e_gamma_l1}
\item $\norm{\hat{\gamma}-\gamma_0}_2=O_{P_n}(\sqrt{s_\gamma\log(p \lor n)/n})$. \label{e_gamma_l2}
\item $\norm{\hat{\beta}(\psi_0)-\beta_0}_1=O_{P_n}(s^*\sqrt{\log(p \lor n)/n})$. \label{e_beta_l1}
\item $\norm{\hat{\beta}(\psi_0)-\beta_0}_2=O_{P_n}(\sqrt{s^*\log(p \lor n)/n})$. \label{e_beta_l2}
\end{enumerate}
where $s^*=s_\gamma\lor s_\beta$. 
\end{assumption}
\begin{assumption}\label{cons_p}
(Rates of convergence for the predictions). For a given sequence $P_n$ we have that
\begin{enumerate}[label=(\roman*)]
\item $\mathbb{E}_{n}[\{\pi(L_i;\gamma_0)-\pi(L_i;\hat{\gamma})\}^2]=O_{P_n}(s_\gamma\log(p \lor n)/n).$ \label{p_pi}
\item $\mathbb{E}_{n}\left([\{m(L_i;\beta_0)-m\{L_i;\hat{\beta}(\psi_0)\}]^2\right)=O_{P_n}(s^*\log(p \lor n)/n)$. \label{p_m}
\end{enumerate}
\end{assumption}
\begin{assumption}\label{strong_cons_CI}
(Dependency on estimated weights). Assuming that $m(L;\beta_0)$ is linear in $\beta_0$, then for a given sequence $P_n$ we have that
\begin{align*}
\norm{\hat{\beta}(\psi_0,\gamma_0)-\hat{\beta}(\psi_0,\hat{\gamma})}_1=O_{P_n}\left( \max_{i\leq n}|H_i(\psi_0)-m(L_i;\beta_0)|\sqrt{\frac{s_\gamma s^* \log (p\lor n)}{n}}\right).
\end{align*}
\end{assumption}
\begin{assumption}\label{bounded_res_CI}
(Regularity conditions on the errors).\\ $\max_{i\leq n}|H_i(\psi_0)-m(L_i;\beta_0)|\sqrt{s_\gamma s^*}\log(p\lor n)=o(\sqrt{n})$ with probability approaching 1.
\end{assumption}

Assumption \ref{moment_CI} places mild moment conditions on the residuals. Given that we are working under model $\mathcal{M}\cap\mathcal{A}$, Assumption \ref{concentration} can be shown to hold using the moderate deviations theory of self-normalised sums \citep{de_la_pena_self-normalized_2009,belloni_sparse_2012}, assuming that $\log (p)=o(n^{1/3})$ and (for example) the regressors $L$ are Gaussian or have bounded support. Alternatively, one can place sub-exponential type conditions on $\mathbb{E}_n[\partial U_i(\psi_0,\eta_0)/\partial \beta]$ as in \citet{ning_general_2017}. The rates in Assumption \ref{cons_e} and  \ref{cons_p} are known to hold for several sparse estimators, including Lasso, post-Lasso and weighted Lasso \citep{belloni_inference_2014,belloni_post-selection_2016}. That rate \ref{cons_p}\ref{p_pi} holds for Lasso logistic regression follows from \citet{farrell_robust_2015}; rate \ref{cons_p}\ref{p_m} holds for weighted lasso and post-lasso estimators \citep{belloni_post-selection_2016}. Note that obtaining these results implies certain restrictions on the penalties, namely that 
\begin{align}
\lambda_\gamma&=O(\sqrt{\log (p\lor n) /n}) \label{pen_ord_gamma}\\
\lambda_\beta&=O(\sqrt{\log (p\lor n)  /n}). \label{pen_ord_beta}
\end{align} 
The result in Assumption \ref{strong_cons_CI} is shown to hold for the proposed estimators of $\beta_0$ in \citet{dukes_high-dimensional_2019} (so long as the model is linear); we refer to that paper for a list of primitive conditions required for it to hold. Assumption \ref{bounded_res_CI} allows us to trade off restrictions on the distribution of the errors with stronger sparsity conditions.   

\subsection*{Proof of Theorem \ref{theorem1}}

\begin{proof}
In order to obtain result (\ref{UC_result}), we will show each of the following in turn:
\begin{align}
&\frac{1}{\sqrt{n}}\sum^n_{i=1}U_i\{\psi_0,\hat{\eta}(\psi_0)\}=\frac{1}{\sqrt{n}}\sum^n_{i=1}U_i(\psi_0,\eta_0)+o_{P_n}(1) \label{S1}\\
&V^{-1/2}\frac{1}{\sqrt{n}}\sum^n_{i=1}U_i(\psi_0,\eta_0)\overset{d}{\to}\mathcal{N}(0,1) \label{S2}\\
&\hat{V}^{-1/2}=V^{-1/2}+o_{P_n}(1). \label{S3}
\end{align}

\textit{Step 1} (Asymptotic linearity). We have
\begin{align*}
&\frac{1}{\sqrt{n}}\sum^n_{i=1}U_i\{\psi_0,\hat{\eta}(\psi_0)\}-\frac{1}{\sqrt{n}}\sum^n_{i=1}U_i(\psi_0,\eta_0)\\
&=\frac{1}{\sqrt{n}}\sum^n_{i=1}\{A_i-\pi(L_i;\gamma_0)\}[m(L_i;\beta_0)-m\{L_i;\hat{\beta}(\psi_0)\}]
\\&\quad+\frac{1}{\sqrt{n}}\sum^n_{i=1}\{H_i(\psi_0)-m(L_i;\beta_0)\}\{\pi(L_i;\gamma_0)-\pi(L_i;\hat{\gamma})\}
\\&\quad+\frac{1}{\sqrt{n}}\sum^n_{i=1}[m\{L_i;\hat{\beta}(\psi_0)\}-m(L_i;\beta_0)]\{\pi(L_i;\hat{\gamma})-\pi(L_i;\gamma_0)\}
\\&=\mathcal{I}_1+\mathcal{I}_2+\mathcal{I}_3.
\end{align*}
 For $\mathcal{I}_1$, using a Taylor expansion,
\begin{align*}
\mathcal{I}_1&=-\frac{1}{\sqrt{n}}\sum^n_{i=1}\bigg\{\frac{\partial U_i(\psi_0,\eta_0)}{\partial \beta}\bigg\} \{\beta_0-\hat{\beta}(\psi_0)\}+O_{P_n}(\sqrt{n}||\beta_0-\hat{\beta}(\psi_0)||^2_2).
\end{align*}
By Assumption \ref{cons_e}\ref{e_beta_l2} and the sparsity condition \ref{US_sum_CI} stated in Theorem \ref{theorem1}, $O_{P_n}(\sqrt{n}||\beta_0-\hat{\beta}(\psi_0)||^2_2)=o_{P_n}(1)$. Then using H\"older's inequality, 
\begin{align*}
&\bigg|\frac{1}{\sqrt{n}}\sum^n_{i=1}\bigg\{\frac{\partial U_i(\psi_0,\eta_0)}{\partial \beta}\bigg\}\{\beta_0-\hat{\beta}(\psi_0)\}\bigg|\\
&\leq\norm{\frac{1}{\sqrt{n}}\sum^n_{i=1}\{A_i-\pi(L_i;\gamma_0)\}d(L_i;\beta_0)}_\infty\norm{\beta_0-\hat{\beta}(\psi_0)}_1\\
&=O_{P_n}(\sqrt{\log(p\lor n)})\norm{\beta_0-\hat{\beta}(\psi_0)}_1
\end{align*}
following Assumption \ref{concentration}. Then by Assumption \ref{cons_e}\ref{e_beta_l1} and condition \ref{US_sum_CI}, $|\mathcal{I}_1|=o_{P_n}(1)$. 

Considering now $\mathcal{I}_2$,
\begin{align*}
&\frac{1}{\sqrt{n}}\sum^n_{i=1}\{H_i(\psi_0)-m(L_i;\beta_0)\}\{\pi(L_i;\gamma_0)-\pi(L_i;\hat{\gamma})\}\\
&=\frac{1}{\sqrt{n}}\sum^n_{i=1}[H_i(\psi_0)-m\{L_i;\hat{\beta}(\psi_0)\}]\{\pi(L_i;\gamma_0)-\pi(L_i;\hat{\gamma})\}\\
&\quad+\frac{1}{\sqrt{n}}\sum^n_{i=1}[m\{L_i;\hat{\beta}(\psi_0)\}-m(L_i;\beta_0)]\{\pi(L_i;\gamma_0)-\pi(L_i;\hat{\gamma})\}\\
&=\mathcal{I}_{2a}+\mathcal{I}_{2b}
\end{align*}
For $\mathcal{I}_{2a}$, by a Taylor expansion,
\begin{align*}
&\frac{1}{\sqrt{n}}\sum^n_{i=1}[H_i(\psi_0)-m\{L_i;\hat{\beta}(\psi_0)\}]\{\pi(L_i;\gamma_0)-\pi(L_i;\hat{\gamma})\}\\
&=-\frac{1}{\sqrt{n}}\sum^n_{i=1}\bigg[\frac{\partial U_i\{\psi_0,\hat{\eta}(\psi_0)\}}{\partial \gamma}\bigg]  (\gamma_0-\hat{\gamma})
+O_{P_n}(\sqrt{n}||\gamma_0-\hat{\gamma}||^2_2)
\end{align*}
and 
\begin{align*}
&\bigg|\frac{1}{\sqrt{n}}\sum^n_{i=1}\bigg[\frac{\partial U_i\{\psi_0,\hat{\eta}(\psi_0)\}}{\partial \gamma}\bigg]  (\gamma_0-\hat{\gamma})\bigg|\\
&\leq \sqrt{n}\norm{\frac{1}{n}\sum^n_{i=1}\frac{\partial U_i\{\psi_0,\hat{\eta}(\psi_0)\}}{\partial \gamma}}_{\infty}||\gamma_0-\hat{\gamma}||_1\\
&\leq\sqrt{n}\lambda_\beta\delta||\gamma_0-\hat{\gamma}||_1
\end{align*}
since $\norm{\delta|\hat{\gamma}|^{\delta-1}\circ\sign(\hat{\gamma})}_\infty \leq 1$ for $\delta\to 1 +$. Then given Assumption \ref{cons_e}\ref{e_gamma_l1}, \ref{cons_e}\ref{e_gamma_l2}, (\ref{pen_ord_gamma}) and condition \ref{US_sum_CI}, it follows that $|\mathcal{I}_{2a}|=o_{P_n}(1)$. Moving onto $\mathcal{I}_{2b}$, by H\"older's inequality
\begin{align*}
&\bigg|\frac{1}{\sqrt{n}}\sum^n_{i=1}[m\{L_i;\hat{\beta}(\psi_0)\}-m(L_i;\beta_0)]\{\pi(L_i;\gamma_0)-\pi(L_i;\hat{\gamma})\}\bigg|\\
&\leq \sqrt{n}\mathbb{E}_n\left([m\{L_i;\hat{\beta}(\psi_0)\}-m(L_i;\beta_0)]^2\right)^{1/2}\mathbb{E}_n[\{\pi(L_i;\gamma_0)-\pi(L_i;\hat{\gamma})\}^2]^{1/2}
\end{align*}
Then given the joint sparsity condition \ref{US_sum_CI} on $s_\gamma$ and $s_\beta$, it follows that
\begin{align*}
&\sqrt{n}\mathbb{E}_n\left([m\{L_i;\hat{\beta}(\psi_0)\}-m(L_i;\beta_0)]^2\right)^{1/2}\mathbb{E}_n[\{\pi(L_i;\gamma_0)-\pi(L_i;\hat{\gamma})\}^2]^{1/2}\\&=o_{P_n}(1)
\end{align*}
Repeating the above reasoning, we have $|\mathcal{I}_3|=o_{P_n}(1)$, and the result (\ref{S1}) follows.


\textit{Step 2} (Asymptotic normality). It follows from Assumptions \ref{moment_CI}\ref{m2} and \ref{moment_CI}\ref{m3} that $\mathbb{E}_{P_n}\{U_i(\psi_0,\eta_0)\}$ is bounded away from zero and above uniformly in $n$. Furthermore, $\mathbb{E}_{P_n}\{|U_i(\psi_0,\eta_0)|^{2+\epsilon}\}\leq C$ by Assumption \ref{moment_CI}\ref{m2}. Hence the Lyapunov condition is verified, and one can invoke the Lyapunov central limit theorem for triangular arrays to arrive at (\ref{S2}).

\textit{Step 3} (Consistency of the variance estimator). Further details on the following arguments can be found in the appendix of \citet{dukes_high-dimensional_2019}. Given that $\mathbb{E}_{P_n}\{U_i(\psi_0,\eta_0)^2\}$ can be bounded  above and below uniformly in $n$ by Assumption \ref{moment_CI}, it will suffice to prove that $\mathbb{E}_{n}[U_i\{\psi_0,\hat{\eta}(\psi_0)]^2\}=\mathbb{E}_{P_n}\{U_i(\psi_0,\eta_0)^2\}+o_{P_n}(1)$. 
One can show that 
\[\mathbb{E}_{n}\{U_i(\psi_0,\eta_0)^2\}=\mathbb{E}_{P_n}\{U_i(\psi_0,\eta_0)^2\}+o_{P_n}(1)\]
using the Von-Bahr Esseen Inequality \citep{von_bahr_inequalities_1965} in combination with Assumptions \ref{moment_CI}\ref{m1} and \ref{moment_CI}\ref{m2}. Then it remains to show that $\mathbb{E}_{n}[U_i\{\psi_0,\hat{\eta}(\psi_0)\}^2]=\mathbb{E}_{n}\{U_i(\psi_0,\eta_0)^2\}=o_{P_n}(1)$. 

Applying the triangle inequality, 
 \begin{align*}
&|\mathbb{E}_{n}[U_i\{\psi_0,\hat{\eta}(\psi_0)\}^2]-\mathbb{E}_{n}\{U_i(\psi_0,\eta_0)^2\}|\\
&\leq \mathbb{E}_{n}[\{\pi(L_i;\hat{\gamma})-\pi(L_i;\gamma_0)\}^2\{H_i(\psi_0)-m(L_i;\beta_0)\}^2]
\\& \quad+ |2\mathbb{E}_{n}[\{A_i-\pi(L_i;\gamma_0)\}\{\pi(L_i;\hat{\gamma})-\pi(L_i;\gamma_0)\}\{H_i(\psi_0)-m(L_i;\beta_0)\}^2]|\\
&\quad+\mathbb{E}_{n}\left([m\{L_i;\hat{\beta}(\psi_0)\}-m(L_i;\beta_0)]^2\{A_i-\pi(L_i;\hat{\gamma})\}^2 \right)\\
&\quad+|2\mathbb{E}_{n}[\{H_i(\psi_0)-m(L_i;\beta_0)\}[m\{L_i;\hat{\beta}(\psi_0)\}-m(L_i;\beta_0)]\{A_i-\pi(L_i;\hat{\gamma})\}^2]|\\
&=\mathcal{I}_4+\mathcal{I}_5+\mathcal{I}_6+\mathcal{I}_7
\end{align*}
Then applying the von Bahr-Esseen inequality and given Assumptions \ref{moment_CI}\ref{m2}, \ref{cons_p}\ref{p_pi}, \ref{cons_p}\ref{p_m} and condition \ref{US_sum_CI}, we have:
\begin{align*}
\mathcal{I}_{4}\leq& \max_{i\leq n} |\pi(L_i;\hat{\gamma})-\pi(L_i;\gamma_0)|\mathbb{E}_{n}[\{\pi(L_i;\hat{\gamma})-\pi(L_i;\gamma_0)\}^2]^{1/2}\\&
\times\mathbb{E}_{n}\{|H_i(\psi_0)-m(L_i;\beta_0)|^4\}^{1/2}=o_{P_n}(1)\\
\mathcal{I}_{5}\leq &2\max_{i\leq n}|A_i-\pi(L_i;\gamma_0)|\mathbb{E}_{n}[\{\pi(L_i;\hat{\gamma})-\pi(L_i;\gamma_0)\}^2]^{1/2}\\&\times \mathbb{E}_{n}\{|H_i(\psi_0)-m(L_i;\beta_0)|^4\}^{1/2}=o_{P_n}(1)\\
\mathcal{I}_{6}\leq& \max_{i\leq n}\{A_i-\pi(L_i;\hat{\gamma})\}^2\mathbb{E}_{n}\left([m\{L_i;\hat{\beta}(\psi_0)\}-m(L_i;\beta_0)]^2\right)=o_{P_n}(1)\\
\mathcal{I}_{7} \leq &2 \max_{i\leq n}\{A_i-\pi(L_i;\hat{\gamma})\}^2\mathbb{E}_{n}[\{H_i(\psi_0)-m(L_i;\beta_0)\}^2]^{1/2}\\&\times\mathbb{E}_{n}\left([m\{L_i;\hat{\beta}(\psi_0)\}-m(L_i;\beta_0)]^2\right)^{1/2}=o_{P_n}(1).
\end{align*}

\textit{Step 4} (Uniform validity). Note that it is immediate from steps 1-3 that
\begin{align*}
T_n\{\psi_0,\hat{\eta}(\psi_0)\}\overset{d}{\to}\mathcal{N}(0,1).
\end{align*}
under any sequence $P_n$. 
Then along the lines in of the proof of Proposition 1 in \citet{chernozhukov_valid_2015}, for any sequence $\delta_n\to0$, let us consider an arbitrary sequence $P^*_n$  where
\begin{align*}\sup_{P_n\in\mathcal{P}'}|\mathbb{P}_{P_n}\left(\psi_0\in[l_s,u_s]\right)-(1-\alpha)|\leq|\mathbb{P}_{P^*_n}\left(\psi_0\in[l_s,u_s]\right)-(1-\alpha)|+\delta_n
\end{align*}
However, by (\ref{S1}), (\ref{S2}) and (\ref{S3}) we have that 
\[\mathbb{P}_{P^*_n}\left(\psi_0\in[l_s,u_s]\right)=\mathbb{P}_{P^*_n}\left[|T_n\{ \psi_0,\hat{\eta}(\psi_0)\}|\leq\Phi(1-\alpha/2)\right]\to1-\alpha\]
and the main result follows. 

\end{proof}

\subsection*{Proof of Corollary \ref{corr1}}

\begin{proof}
We begin by noting that one can motivate the proposed estimator of $\gamma$ as the solution to the penalised estimating equations
\begin{align*}
0&=\frac{1}{n}\sum^n_{i=1}\frac{\partial}{\partial \beta}U_i\{\psi_0,\hat{\eta}(\psi_0)\}+\lambda_\gamma\delta|\hat{\gamma}|^{\delta-1}\circ\sign(\hat{\gamma})\nonumber\\
&=\frac{1}{n}\sum^n_{i=1}-\{A_i-\pi(L_i;\hat{\gamma})\}L_i+\lambda_\gamma\delta|\hat{\gamma}|^{\delta-1}\circ\sign(\hat{\gamma}),
\end{align*}
letting $\delta\to1 +$. 

Then repeating the arguments in Step 1 of Theorem 1, for $\mathcal{I}_1$ we now have
\begin{align}
\mathcal{I}_1&=\frac{1}{\sqrt{n}}\sum^n_{i=1}\{A_i-\pi(L_i;\hat{\gamma})\}[m(L_i;\beta_0)-m\{L_i;\hat{\beta}(\psi_0)\}]\nonumber\\
&\quad+\frac{1}{\sqrt{n}}\sum^n_{i=1}\{\pi(L_i;\hat{\gamma})-\pi(L_i;\gamma_0)\}[m(L_i;\beta_0)-m\{L_i;\hat{\beta}(\psi_0)\}]
\end{align}
The second term in the right hand side is $o_{P_n}(1)$ under Assumption \ref{cons_p} and sparsity condition \ref{US_sum_CI}. Then
\begin{align*}
&\frac{1}{\sqrt{n}}\sum^n_{i=1}\{A_i-\pi(L_i;\hat{\gamma})\}[m(L_i;\beta_0)-m\{L_i;\hat{\beta}(\psi_0)\}]\\&=-\frac{1}{\sqrt{n}}\sum^n_{i=1}\left\{\frac{\partial U_i\{\psi_0,\hat{\eta}(\psi_0)\}}{\partial \beta}\right\}\{\beta_0-\hat{\beta}(\psi_0)\}+O_{P_n}(\sqrt{n}||\beta_0-\hat{\beta}(\psi_0)||^2_2).
\end{align*}
and 
\begin{align*}
&\bigg|\frac{1}{\sqrt{n}}\sum^n_{i=1}\left\{\frac{\partial U_i\{\psi_0,\hat{\eta}(\psi_0)\}}{\partial \beta}\right\}\{\beta_0-\hat{\beta}(\psi_0)\}\bigg|\\&\leq\sqrt{n}\norm{\frac{1}{n}\sum^n_{i=1}\frac{\partial U_i\{\psi_0,\hat{\eta}(\psi_0)\}}{\partial \beta}}_\infty\norm{\beta_0-\hat{\beta}(\psi_0)}_1\\
&\leq\sqrt{n}\lambda_\gamma\delta||\beta_0-\hat{\beta}(\psi_0)||_1
\end{align*}
since $\norm{\delta|\hat{\beta}(\psi_0)|^{\delta-1}\circ\sign\{\hat{\beta}(\psi_0)\}}_\infty \leq 1$ for $\delta\to 1 +$. Then by Assumptions \ref{cons_e}\ref{e_beta_l1}, \ref{cons_e}\ref{e_beta_l2}, (\ref{pen_ord_beta}) and condition \ref{US_sum_CI}, $|\mathcal{I}_{1b}|=o_{P_n}(1)$. The result follows by repeating the remaining steps in the above proof.
\end{proof}

\subsection*{Proof of Theorem \ref{theorem2}}
\noindent\textbf{Proof for linear models}

\begin{proof}
We first consider the case where $m(L;\beta_0)$ is linear in $\beta_0$  (so that we can invoke Assumption \ref{strong_cons_CI}). Repeating Step 1 of the proof of Theorem 1, for $\mathcal{I}_1$ we now have 
\begin{align*}
&\frac{1}{\sqrt{n}}\sum^n_{i=1}\{A_i-\pi(L_i;\gamma_0)\}\{m(L_i;\beta_0)-m(L_i;\hat{\beta}(\psi_0,\hat{\gamma}))\}\\
&=\frac{1}{\sqrt{n}}\sum^n_{i=1}\{A_i-\pi(L_i;\gamma_0)\}\{m(L_i;\beta_0)-m\{L_i;\hat{\beta}(\psi_0,\gamma_0)\}\}\\
&\quad+\frac{1}{\sqrt{n}}\sum^n_{i=1}\{A_i-\pi(L_i;\gamma_0)\}\{m\{L_i;\hat{\beta}(\psi_0,\gamma_0)\}-m(L_i;\hat{\beta}(\psi_0,\hat{\gamma}))\}\\
&=\mathcal{I}_{1a}+\mathcal{I}_{1b}
\end{align*}

Considering first $\mathcal{I}_{1a}$, by the location-shift condition \ref{l_shift} in Theorem 2,
\begin{align*}
&\mathbb{E}_{P_n}[\mathcal{I}_{1a}|\{H_i(\psi_0),L_i\}^n_{i=1}]\\&=\frac{1}{\sqrt{n}}\sum^n_{i=1}\left(\mathbb{E}_{P_n}[A_i|\{H_i(\psi_0),L_i\}^n_{i=1}]-\pi(L_i;\gamma_0)\right)[m(L_i;\beta_0)-m\{L_i;\hat{\beta}(\psi_0,\gamma_0)]\\
&=0
\end{align*}
and 
\begin{align*}
&\mathbb{E}_{P_n}[\mathcal{I}^{2}_{1a}|\{H_i(\psi_0),L_i\}^n_{i=1}]
\leq C\mathbb{E}_{n}\left([m(L_i;\beta_0)-m\{L_i;\hat{\beta}(\psi_0,\gamma_0)\}]^2\right).
\end{align*}
By Assumption \ref{cons_p}\ref{p_m} and sparsity condition \ref{S_sum_CI}, $\mathbb{E}_{P_n}[R^2_{1a}]=o(1)$, and therefore$|R_{1a}|=o_{P_n}(1)$.

For $\mathcal{I}_{1b}$,
\begin{align*}
&\bigg|\frac{1}{\sqrt{n}}\sum^n_{i=1}\{A_i-\pi(L_i;\gamma_0)\}[m\{L_i;\hat{\beta}(\psi_0,\gamma_0)\}-m\{L_i;\hat{\beta}(\psi_0,\hat{\gamma})\}]\bigg|&\nonumber\\
&\leq \norm{\frac{1}{\sqrt{n}}\sum^n_{i=1}\{A_i-\pi(L_i;\gamma_0)\}L_i }_\infty\norm{\hat{\beta}(\psi_0,\gamma_0)-\hat{\beta}(\psi_0,\hat{\gamma})}_1\\
&=O_{P_n}(\sqrt{\log (p\lor n)})\norm{\hat{\beta}(\psi_0,\gamma_0)-\hat{\beta}(\psi_0,\hat{\gamma})}_1=o_{P_n}(1)
\end{align*}
by Assumptions \ref{strong_cons_CI}, \ref{bounded_res_CI} and sparsity conditions \ref{S_sum_CI} and  \ref{prod_CI}. Note that if $\beta_0$ is estimated without weights, this step is not required.

Then for $\mathcal{I}_{2}$, 
\begin{align*}
&\mathbb{E}_{P_n}\{\mathcal{I}_{2}|(A_i,L_i)^n_{i=1}\}\\&=\mathbb{E}_{n}\left(\mathbb{E}_{P_n}[\{H_i(\psi_0)-m(L;\beta_0)\}^2|(A_i,L_i)^n_{i=1}]\{\pi(L_i;\gamma_0)-\pi(L_i;\hat{\gamma})\}^2\right)\\&\leq C\mathbb{E}_{n}[\{\pi(L_i;\gamma_0)-\pi(L_i;\hat{\gamma})\}^2],
\end{align*}
by Assumption \ref{moment_CI}\ref{m1}; then $|\mathcal{I}_{2}|=o_{P_n}(1)$ given condition \ref{S_sum_CI}. One can show that $|\mathcal{I}_{3}|=o_{P_n}(1)$ under Assumption \ref{cons_p} and sparsity condition \ref{prod_CI}, without requiring the condition \ref{US_sum_CI} that was invoked in the proof of Theorem \ref{theorem1}. Indeed, one can then repeat steps 2-4 of the proof of Theorem \ref{theorem1}, invoking conditions \ref{S_sum_CI} and \ref{prod_CI} instead of \ref{US_sum_CI}, to obtain the main result.
\end{proof}

\noindent\textbf{Proof using sample splitting}

\begin{proof}
For non-linear models, for simplicity we will consider a simple scheme whereby the data is split into approximately equal subsamples $k$ and $k^c$, where $k$ has sample size $n_k=n/2$ (so $n_{k^c}=n-n_k$). This sketch proof also extends to more complex schemes as in \citet{chernozhukov_double/debiased_2018}. We estimate $\beta_0$ using sample $k^c$ only, such that the resulting estimates will be denoted by $\hat{\beta}^{k^c}(\psi_0)$. Let $\hat{\gamma}^{k}$ denote an estimate of $\gamma_0$ obtained using sample $k$ (sample $k^c$ could be used instead without changing the final result). Then 
\begin{align*}
&\frac{1}{\sqrt{n_k}}\sum^{n_k}_{i=1}U^k_i(\psi_0,\hat{\eta}^{k^c})-\frac{1}{\sqrt{n_k}}\sum^{n_k}_{i=1}U_i(\psi_0,\eta_0)\\
&=\frac{1}{\sqrt{n_k}}\sum^{n_k}_{i=1}\{A^k_i-\pi(L^k_i;\gamma_0)\}[m(L^k_i;\beta_0)-m\{L^k_i;\hat{\beta}^{k^c}(\psi_0)\}]
\\&\quad+\frac{1}{\sqrt{n_k}}\sum^{n_k}_{i=1}\{H^k_i(\psi_0)-m(L^k_i;\beta_0)\}\{\pi(L^k_i;\gamma_0)-\pi(L^k_i;\hat{\gamma}^{k})\}
\\&\quad+\frac{1}{\sqrt{n_k}}\sum^{n_k}_{i=1}[m\{L^k_i;\hat{\beta}^{k^c}(\psi_0)\}-m(L^k_i;\beta_0)]\{\pi(L^k_i;\hat{\gamma}^{k})-\pi(L^k_i;\gamma_0)\}.
\\&=\check{\mathcal{I}}_1+\check{\mathcal{I}}_2+\check{\mathcal{I}}_3.
\end{align*}
By Assumptions \ref{moment_CI}\ref{m1}, 
\begin{align*}
\mathbb{E}_{P_n}\{\check{\mathcal{I}}^2_1|(L^k_i)^{n_k}_{i=1},k^c\}&\leq C\mathbb{E}_{n}\left([m(L^k_i;\beta_0)-m\{L^k_i;\hat{\beta}^{k^c}(\psi_0)\}]^2\right)\\
\mathbb{E}_{P_n}\{\check{\mathcal{I}}^2_2|(A^k_i, L^k_i)^{n_k}_{i=1}\}&\leq C\mathbb{E}_{n}[\{\pi(L^k_i;\gamma_0)-\pi(L^k_i;\hat{\gamma}^{k})\}^2]
\end{align*}
and 
\begin{align*}
|\check{\mathcal{I}}_3|\leq \sqrt{n}\mathbb{E}_{n}\left([m\{L^k_i;\hat{\beta}^{k^c}(\psi_0)\}-m(L^k_i;\beta_0)]^2\right)^{1/2}\mathbb{E}_{n}[\{\pi(L^k_i;\hat{\gamma}^{k})-\pi(L^k_i;\gamma_0)\}^2]^{1/2}
\end{align*}
Hence invoking Assumptions \ref{cons_p} and conditions \ref{S_sum_CI} and \ref{prod_CI}, it follows that $ \check{\mathcal{I}}_1$, $\check{\mathcal{I}}_2$ and $\check{\mathcal{I}}_3$ are all $o_{P_1}(1)$. Validity of the confidence intervals follows from repeating steps 2-4 of the proof of Theorem \ref{theorem1}.
\end{proof}

\newpage 
\section{Appendix B}\label{appB} 
\subsection{Additional simulation results}

\begin{table}[htbp]
\centering
\caption{Simulation results from repeating Experiments 1-3 at $n=p=400$. Estimators considered (Est); Monte Carlo bias multiplied by 10 (Bias); Monte Carlo standard deviation multiplied by 10 (MCSD); Mean standard error multiplied by 10 (MSE); coverage probability multiplied by 100 (Cov).}
\resizebox{\textwidth}{!}{\begin{tabular}{llllllllllllll}
\hline
   &  & \multicolumn{4}{c}{Experiment 1} & \multicolumn{4}{c}{Experiment 2} & \multicolumn{4}{c}{Experiment 3}\\
 \textbf{$\rho,\tau$} & Est & Bias & MCSD & MSE & Cov & Bias & MCSD & MSE & Cov & Bias & MCSD & MSE & Cov\\
\hline
2,1 & $\hat{\psi}_{OLS}$ & -0$\cdot$54 & 1$\cdot$4 & 1$\cdot$1 & 86$\cdot$2 & -2$\cdot$94 & 2$\cdot$7 & 2 & 63$\cdot$6 & -0$\cdot$59 & 1$\cdot$2 & 1$\cdot$1 & 88$\cdot$3 \\ 
   & $\hat{\psi}_{PDS}$ & -0$\cdot$39 & 1$\cdot$2 & 1$\cdot$2 & 93 & -3$\cdot$74 & 2$\cdot$1 & 2$\cdot$1 & 57$\cdot$6 & -0$\cdot$42 & 1$\cdot$1 & 1$\cdot$1 & 92$\cdot$5 \\ 
   & $\hat{\psi}_{PO}$ & -0$\cdot$62 & 1$\cdot$1 & 1$\cdot$1 & 91$\cdot$7 & -3$\cdot$74 & 2$\cdot$1 & 2$\cdot$2 & 62$\cdot$1 & -0$\cdot$65 & 1 & 1$\cdot$1 & 93$\cdot$2 \\ 
   & $\hat{\psi}_{PDS-CV}$ & -0$\cdot$56 & 1$\cdot$3 & 1$\cdot$2 & 91$\cdot$4 & -3$\cdot$63 & 2$\cdot$2 & 2$\cdot$2 & 61$\cdot$7 & -0$\cdot$64 & 1$\cdot$2 & 1$\cdot$2 & 91$\cdot$9 \\ 
   & $\hat{\psi}_{PO-CV}$ & -0$\cdot$66 & 1$\cdot$3 & 1$\cdot$2 & 89$\cdot$4 & -3$\cdot$63 & 2$\cdot$2 & 2$\cdot$1 & 59$\cdot$8 & -0$\cdot$77 & 1$\cdot$1 & 1$\cdot$2 & 89$\cdot$2 \\ 
   & $\hat{\psi}_{HDBR}$ & -0$\cdot$34 & 1$\cdot$4 & 1$\cdot$4 & 94$\cdot$6 & 0$\cdot$14 & 2 & 2 & 95$\cdot$6 & -0$\cdot$36 & 1$\cdot$2 & 1$\cdot$2 & 93$\cdot$6 \\ 
  0$\cdot$5,0$\cdot$4 & $\hat{\psi}_{OLS}$ & -1$\cdot$42 & 1$\cdot$8 & 1$\cdot$2 & 68$\cdot$3 & -0$\cdot$95 & 2$\cdot$2 & 1$\cdot$8 & 87$\cdot$2 & -1$\cdot$4 & 1$\cdot$8 & 1$\cdot$2 & 67$\cdot$6 \\ 
   & $\hat{\psi}_{PDS}$ & -2$\cdot$2 & 1$\cdot$5 & 1$\cdot$3 & 59$\cdot$3 & -2$\cdot$75 & 2 & 2$\cdot$1 & 74$\cdot$8 & -2$\cdot$3 & 1$\cdot$5 & 1$\cdot$3 & 54$\cdot$4 \\ 
   & $\hat{\psi}_{PO}$ & -2$\cdot$22 & 1$\cdot$5 & 1$\cdot$3 & 58$\cdot$8 & -2$\cdot$75 & 2 & 2$\cdot$2 & 78$\cdot$3 & -2$\cdot$32 & 1$\cdot$5 & 1$\cdot$3 & 55$\cdot$6 \\ 
   & $\hat{\psi}_{PDS-CV}$ & -0$\cdot$78 & 1$\cdot$4 & 1$\cdot$4 & 90 & -2$\cdot$79 & 2$\cdot$1 & 2$\cdot$2 & 77$\cdot$3 & -0$\cdot$72 & 1$\cdot$4 & 1$\cdot$4 & 91$\cdot$2 \\ 
   & $\hat{\psi}_{PO-CV}$ & -0$\cdot$92 & 1$\cdot$3 & 1$\cdot$3 & 87$\cdot$1 & -2$\cdot$8 & 2 & 2$\cdot$2 & 75$\cdot$9 & -0$\cdot$85 & 1$\cdot$4 & 1$\cdot$3 & 86$\cdot$1 \\ 
   & $\hat{\psi}_{HDBR}$ & -0$\cdot$56 & 1$\cdot$5 & 1$\cdot$6 & 94$\cdot$4 & 0$\cdot$07 & 2$\cdot$2 & 2$\cdot$2 & 95$\cdot$5 & -0$\cdot$55 & 1$\cdot$5 & 1$\cdot$6 & 93$\cdot$8 \\ 
\hline
\end{tabular}}
\end{table}

\bibliographystyle{apalike}
\bibliography{HDinf_MM}

\end{document}